\documentclass[journal=jpcbfk,manuscript=article]{achemso}
\setkeys{acs}{articletitle = true}
\usepackage[version=3]{mhchem} 
\usepackage{color}
\usepackage{amsfonts}
\usepackage{amssymb}
\usepackage{amsmath}
\usepackage{graphicx}
\usepackage{float}

\author{Nicolas Bergmann}
\affiliation{Department of Chemistry, Technical University of Munich, D-85748 Garching, Germany}

\author{Michael Galperin}
\email{migalperin@ucsd.edu}
\phone{+1 858 246 0511}
\affiliation{Department of Chemistry \& Biochemistry, University of California San Diego, La Jolla, CA 92093, USA}

\title[Hubbard NEGF]{Electron Transfer Methods in Open Systems}


\begin{document}

\begin{abstract}
Utilization of electron transfer methods for description of quantum transport is
popular due to simplicity of the formulation and its ability to account for basic 
physics of electron exchange between system and baths. 
At the same time, necessity to go beyond simple golden rule-type expressions for
rates was indicated in the literature and {\em ad hoc} formulations were proposed.
Similarly, kinetic schemes for quantum transport beyond usual second order
Lindblad/Redfield considerations were discussed. 
Here we utilize recently introduced by us nonequilibrium Hubbard Green's functions 
diagrammatic technique to analyze construction of rates in open systems.
We show that previous considerations for rates of second and fourth order 
can be obtained as a particular case of zero and second order Green's function
diagrammatic series with bare diagrams. We discuss limitations of 
previous considerations, stress advantages of the Hubbard
Green's function approach in constructing the rates and indicate
that standard dressing of the diagrams is a natural way to account
for additional baths/degrees of freedom when formulating generalized
expressions for the rates.
\end{abstract}

\maketitle
\newpage
\section{Introduction}\label{intro}
Electron transfer processes are the heart of oxidation-reduction reactions which play important role 
in chemistry and biology~\cite{NewtonSutinAnnRevPhysChem84,Kuznetsov_1995,JortnerBixon_1999}.
Theoretical description of electron transfer rates at the level of the Marcus theory~\cite{Marcus1956,Marcus1956a,LevichDogonadze1959,LevichDogonadze1960,HushTransFarSoc61,HushElectrochimActa68,JortnerJCP76}
is widely utilized for the description of variety of phenomena from photovoltaics, batteries design and 
catalysis in chemistry~\cite{MillerCalcaterraClossJACS84,MillerBeitzHuddlestonJACS84,KuharskiBaderChandlerJCP88,BarbaraMeyerRatnerJPC96,RossoDupuisJCP04,Nitzan_2006} 
to photosynthesis, vision and sense of smell in biology~\cite{KukiWolynesScience87,BeratanBettsOnuchicScience91,BeratanOnuchicWinklerGrayScience92,StuchebrukhovJCP96_1,StuchebrukhovJCP96_2,BrookesPRL07,brookes_swipe_2012}.

Interfacial electron transfer is behind many vital biological processes~\cite{mclendon,Miyashita3558,nojiri_direct_2009,Xiong1075,mcgrath_structiral_2015}.
Recently, bio\-mo\-lecu\-les were utilized as building blocks in electric circuits. Biomolecular junctions
are useful as a tool to study properties of molecules and as potential bioelectronic devices. 
For example, electron transport was measured through DNA~\cite{cohen_direct_2005,g_ohler_spin_2011,xie_spin_2011},
oligopeptides~\cite{mondal_photospintronics:_2016,banerjee-ghosh_separation_2018}
and electron transfer proteins~\cite{mentovich_doped_2012,amdursky_electron_2015,yu_insights_2015}.
Also, STM junctions were suggested as a convenient tool for DNA detection and 
sequencing~\cite{zwolak_colloquium_2008}.

Traditionally, electron transfer theory considers isolated donor-bridge-acceptor systems.
Nevertheless, this theoretical approach appears to be useful also in description of 
electron transport in open molecular systems such as, e.g., redox molecular 
junctions~\cite{migliore_nonlinear_2011,migliore_irreversibility_2013,craven_electron_2016,craven_electrothermal_2017}.
Formal relationship between electron transfer rates and molecular conduction was discussed
in the literature~\cite{nitzan_relationship_2001}. 

In both intra-system and interfacial processes in condensed phase electron transfer
may be assisted by intermediates. For example, intermediate states play important role
in photovoltaic~\cite{ALBINSSON2008138} or long-range DNA electron transfer 
processes~\cite{jortner_charge_1998,bixon_long-range_1999,bixon_long-range_2002}.
Effects of intermediates on electron transfer are often discussed in terms of super-exchange
vs. hopping mechanism. Clearly, the two mechanisms are limiting cases (coherent and 
completely incoherent) of the same process, and attempts to unify electron transfer
rate expression were done in the literature~\cite{sumi_unified_2001,saito_unified_2009}.
However, these attempts are {\em ad hoc} perturbation theory considerations,
and an ordered way of rate simulations is still lacking.

Constructing generalized expressions for rates is also at the heart of kinetic equations
approaches  to transport. For example, Refs.~\citenum{LeijnseWegewijsPRB08,koller_density-operator_2010} are similar in  the spirit effort to introduce a scheme for constructing generalized expressions for rates. Advantage of this approach is
possibility to account for higher order processes in system-bath couplings in 
an ordered way of bare perturbation theory. Still, questions of, e.g., accounting
for additional degrees of freedom (such as other baths) within a particular order
in the system-primary baths coupling remains open. Also, the methodology 
has usual for quantum master equations restriction of applicability only in the
high temperature regime (thermal energy should be big relative to characteristic energy
of the system-bath coupling). We note in passing that in interacting open systems kinetic schemes 
should be applied with caution~\cite{nitzan_kinetic_2018}.

Nonequilibrium Green's functions (NEGF)~\cite{HaugJauho_1996} 
is a tool capable to describe both super-exchange
and hopping transport regimes, as well as smooth transition between the two limiting cases. 
Diagrammatic perturbation theory is a way to account for interaction with intermediate 
(or additional) degrees of freedom in an ordered form. Also, Green's function considerations
are applicable in any temperature regime. However, identifying rates is not
possible within the standard NEGF.

Here, we show that recently introduced by us many-body NEGF flavor, 
the Hubbard NEGF~\cite{chen_nonequilibrium_2017}, while retaining advantages 
of the Green's function methods is capable to provide connection with kinetic schemes. 
So that ordered construction of generalized rates becomes possible employing 
the Hubbard Green's function diagrammatic technique.
Structure of the paper is the following. After introducing junction model
we give a short overview of the Hubbard NEGF introducing conceptual 
details not presented in its original introduction in Ref.~\citenum{chen_nonequilibrium_2017}.
After this we present connection between the Hubbard NEGF and kinetic schemes of Refs.~\citenum{LeijnseWegewijsPRB08,koller_density-operator_2010} 
showing a way to formulate rates within the Hubbard NEGF. 
We conclude by discussing advantages of Green's functions formulation and 
outlining directions for future research.

\section{Model}\label{model}
We consider generic model of a junction which consists of a molecule, $M$, coupled to two contacts, 
$L$ and $R$.
Depending on particular problem, molecular (system) part can describe
electronic, vibrational, optically-dressed (e.g. polariton) or any other degrees of freedom.
We assume that quantum chemistry problem for the isolated system has been solved,
and many-body eigenstates $\lvert S\rangle$ and their energies $E_S$ are known.
We note in passing that even in systems with very big (or even infinite) number
of many-body eigenstates, energetics of the junction (bias, driving laser field, etc.)
allows to identify a finite subset, which is enough for first principles simulation
of experimental data for realistic systems~\cite{yeganeh_transport_2009,white_raman_2014,miwa_hubbard_2019,miwa_many-body_2019,kimura_selective_2019}. 
Contacts (baths) are assumed to be  reservoirs of free charge carriers each at its own equilibrium.
Second quantization is utilized to treat baths' degrees of freedom.
This is minimal model for discussion of electron transfer rates within the Hubbard NEGF.
Additional baths and/or degrees of freedom (e.g., phonons) can be added to the consideration in
a straightforward manner.

Hamiltonian of the minimal model is
\begin{align}
\label{H}
 \hat H &= \hat H_M + \sum_{K=L,R}\bigg( \hat H_K + \hat V_{MK}\bigg)
 \\
 \label{HM}
 \hat H_M &= \sum_{S\in M} \lvert S\rangle E_S \langle S\rvert
 \\
 \label{HK}
 \hat H_K &= \sum_{k\in K} \varepsilon_k\hat c_k^\dagger\hat c_k
 \\
 \label{VMK}
 \hat V_{MK} &= \sum_{S_1,S_2 \in M}
 \sum_{k\in K}\bigg( V_{k,S_1S_2}\, \hat c_k^\dagger\, \lvert S_1\rangle\langle S_2\rvert + H.c.\bigg) 
\end{align}
Here $\hat H_M$ and $\hat H_K$ ($K=L,R$) are molecule and contacts Hamiltonians,
and $\hat V_{MK}$ introduces system-baths coupling.
$\hat c_k^\dagger$ ($\hat c_k$) creates (annihilates) electron in single-electron state $k$ 
in the contacts. $V_{k,S_1S_2}$ is matrix element for electron transfer from system to bath
in which system goes from state $\lvert S_2\rangle$ to $\lvert S_1\rangle$.

Below we will be interested in electron flux through the junction and corresponding 
intra-system electron transfer rates. Note that while we focus on charge current,
the consideration is more general and can equivalently be applied to calculation of 
any other intra-system rates (e.g., related to photon or energy flux) or to
multi-time correlation functions as employed in, e.g., nonlinear optical 
spectroscopy~\cite{mukamel_principles_1995}.


\section{Hubbard NEGF}\label{hub}
Central object of interest is the single particle Hubbard Green's function, which is defined 
on the Keldysh contour as
\begin{equation}
\label{defG}
 G_{S_1S_2,S_3S_4}(\tau,\tau') = -i
 \langle T_c\, \hat X_{S_1S_2}(\tau)\,\hat X^\dagger_{S_3S_4}(\tau')\rangle
\end{equation}
Here $X_{S_1S_2}\equiv \lvert S_1\rangle\langle S_2\rvert$ is the Hubbard (or projection)
operator, $T_c$ is the contour ordering operator, and $\tau$ and $\tau'$ are the contour
variables. Advantage of the Hubbard over standard NEGF for our study is 
possibility to access information on many-body states of the system. 
As we show below, this moment is crucial for formulating
general expressions for rates within the Green's function methodology.

Historically, Hubbard Green's functions where introduced for treatment of strongly correlated
extended (lattice type) equilibrium systems in Ref.~\citenum{hubbard_electron_1967}.
Diagrammatic technique (expansion around atomic limit) for such equilibrium Green's functions~\cite{IzyumovSkryabin_1988,OvchinnikovValkov_2004}
is based on assumption of equilibrium character of the uncoupled system's density operator
\begin{equation}
\label{rho0}
\hat \rho_0 = \frac{1}{Z_0}e^{-\hat H_0/k_BT};
\qquad Z_0=\mbox{Tr}\, e^{-\hat H_0/k_BT}
\end{equation}
At nonequilibrium, Hubbard Green's  functions where used for transport simulations
employing relations derived from equation-of-motion considerations and  
functional derivatives in auxiliary fields~\cite{sandalov_theory_2003,FranssonPRB05,SandalovJPCM06,SandalovPRB07,MGNitzanRatnerPRB08,Fransson_2010}.
While the latter approach is very useful, it lacks rigor of ordered diagrammatic expansion
and provides only vague rules about choice of auxiliary fields and terms resulting from performing
functional derivatives.

Recently, we introduced nonequilibrium version of the Hubbard diagrammatic technique~\cite{chen_nonequilibrium_2017} making it applicable to nonequilibrium
impurity-type (molecular junction) problems. Contrary to original lattice type formulation,
we introduce baths and utilize state representation only for the system, while bath degrees of
freedom are treated within standard second quantization.
Thus, perturbative expansion in the system-baths couplings in each order yields
product of two multi-time correlation functions: one for the system and one for the bath operators
(see Ref.~\citenum{chen_nonequilibrium_2017} for details).
Because baths are assumed to be non-interacting, the latter can be treated using 
the standard Wick's theorem. 
To evaluate multi-time correlation function of Hubbard (system) operators we employ
usual for NEGF assumption of steady-state being independent of initial condition
at infinite past. Thus, assuming equilibrium initial system state, Eq.~(\ref{rho0}), correlation function
of Hubbard operators  is evaluated using diagrammatic technique 
of Refs.~\citenum{IzyumovSkryabin_1988, OvchinnikovValkov_2004}.
As a result, Hubbard NEGF appears to be a modified (by presence of baths and nonequilibrium 
character of the system) version of the lattice diagrammatic technique
(see Ref.~\citenum{chen_nonequilibrium_2017} for details).
The technique appears to be quite stable over wide range of parameters~\cite{miwa_towards_2017},
helpful in evaluation of electronic friction in junctions~\cite{chen_current-induced_2019,chen_electronic_2019},
and useful as a convenient tool in first principles simulations of optoelectronic
devices~\cite{miwa_hubbard_2019,miwa_many-body_2019,kimura_selective_2019}.

It is important to realize, however, that requirement of equilibrium character of the uncoupled system 
density operator, Eq.~(\ref{rho0}), in principle can be relaxed. 
Indeed, diagrammatic technique of Refs.~\citenum{IzyumovSkryabin_1988, OvchinnikovValkov_2004} 
is based on commutation properties of the Hubbard operators (interaction representation)
\begin{equation}
 \hat X_{S_1S_2}(t)\,\hat X_{S_3S_4}^\dagger(t') = 
 e^{-i(E_{S_2}-E_{S_1})(t-t')}\delta_{S_2,S_4}\hat X_{S_1S_3}(t')
\pm \hat X_{S_3S_4}^\dagger(t')\,\hat X_{S_1S_2}(t)
\end{equation}
with sign chosen according to the operators statistics (Bose or Fermi - see Ref.~\citenum{chen_nonequilibrium_2017} for details)
and ability to interchange Hubbard and equilibrium density  operators
\begin{equation}
 \hat X_{S_1S_2}\,\hat \rho_0 = \hat \rho_0\, \hat X_{S_1S_2}\, e^{(E_{S_1}-E_{S_2})/k_BT}
\end{equation}
Thus, it is clear that diagrammatic technique for the Hubbard operators
can be equivalently formulated for any form of system's density operator as long 
as the latter is a function of molecular (system) Hamiltonian only
\begin{equation}
\label{rho0_any}
 \hat \rho_0 = f(\hat H_M) \equiv \sum_{S} \lvert S\rangle\, f(E_S)\, \langle S\rvert
\end{equation}
While this observation does not change numerical procedure for the Hubbard NEGF,
it yields two important conceptual consequences.
First, the Hubbard NEGF can be considered as a natural tool for expansion
around results of a quasi-particle-type consideration performed in the basis
of many-body states capable to introduce states broadening and bath-induced
coherences, which were missed in the latter. For example, such quasi-particle-type
consideration is the Markov Redfield/Lindblad quantum master equation. 
In this case $f(E_S)$  in (\ref{rho0_any}) is probability $P_S$ of state $\lvert S\rangle$ to be observed. 
Second, considering situation where only one state is populated, $P_S=1$,
nonequilibrium diagrammatic technique for Hubbard Green's functions
provides access to traditional expressions for rates of transitions from state $\lvert S\rangle$ 
to all other states of the system. Below we discuss details of this 
Green's function-to-kinetic scheme connection.  


\section{Connection to kinetic schemes}\label{kinetic}
We now turn to discuss how Hubbard NEGF can be used to define generalized expressions
for transfer rates. We explore connection to the kinetic scheme presented
in Refs.~\citenum{LeijnseWegewijsPRB08,koller_density-operator_2010},
thus identifying rate expressions in terms of the Hubbard NEGF,
and indicate how the rates expressions can be generalized.

We start by considering equation-of-motion for the probability 
of state $\lvert S\rangle$ to be observed (Heisenberg picture)
\begin{equation}
\label{PS}
 P_S(t)\equiv \langle \hat X_{SS}(t) \rangle
\end{equation} 
under driving by the Hamiltonian $\hat H$ of (\ref{H}). 
Writing Heisenberg equation of motion for (\ref{PS}) and using standard NEGF derivation
(similar to the derivation leading to the celebrated Meir-Wingreen expression 
for current~\cite{HaugJauho_1996})
one gets
\begin{align}
\label{EOM}
\frac{d}{dt} P_S(t) &= 2\mbox{Re}\sum_{S_1,S_2,S_3}\int_{-\infty}^{t} dt'
\nonumber \\
&\bigg(\,\,
 \sigma^{>}_{SS_3,S_1S_2}(t-t')\, G^{<}_{S_1S_2,SS_3}(t'-t)
-\sigma^{<}_{SS_3,S_1S_2}(t-t')\, G^{>}_{S_1S_2,SS_3}(t'-t)
\\
&-\sigma^{>}_{S_3S,S_1S_2}(t-t')\, G^{<}_{S_1S_2,S_3S}(t'-t)
+\sigma^{<}_{S_3S,S_1S_2}(t-t')\, G^{>}_{S_1S_2,S_3S}(t'-t) \bigg)
\nonumber
\end{align}
Here $G^{<(>)}$ is the lesser (greater) projection of the Hubbard Green function (\ref{defG})
and $\sigma^{<(>)}$ is the lesser (greater) projection of the self-energy due to coupling
to contacts $L$ and $R$ 
\begin{equation}
\label{sigma}
 \sigma_{S_1S_2,S_3S_4}(\tau,\tau') = \sum_{K=L,R}\sum_{k\in K} 
 V_{S_1S_2,k}\, g_k(\tau,\tau')\, V_{k,S_3S_4} 
 \equiv \sum_{K=L,R} \sigma^K_{S_1S_2,S_3S_4}(\tau,\tau')
\end{equation}
where $g_k(\tau,\tau')=-\langle T_c\, \hat c_k(\tau)\,\hat c_k^\dagger(\tau')\rangle$
is the Green function of free electron in single-particle state $k$.

Expression (\ref{EOM}) is exact, and our goal is to represent it in the form of rate 
equation, which will allow to identify expressions for the rates.
Following Refs.~\citenum{LeijnseWegewijsPRB08,koller_density-operator_2010}
we will be interested in rates of second and fourth order in the system-baths couplings.
Taking into account that Hubbard NEGF diagrammatic technique expands in
system-baths coupling and noting that second order in the coupling already enters
(\ref{EOM}) via self-energy (\ref{sigma}), it is natural to expect that second order
rates should result from zero order of the Hubbard GFs expansion, while fourth order rates
should be accessible from second order of the Hubbard NEGF diagrammatic series.

\begin{figure}[htbp]
\centering\includegraphics[width=\linewidth]{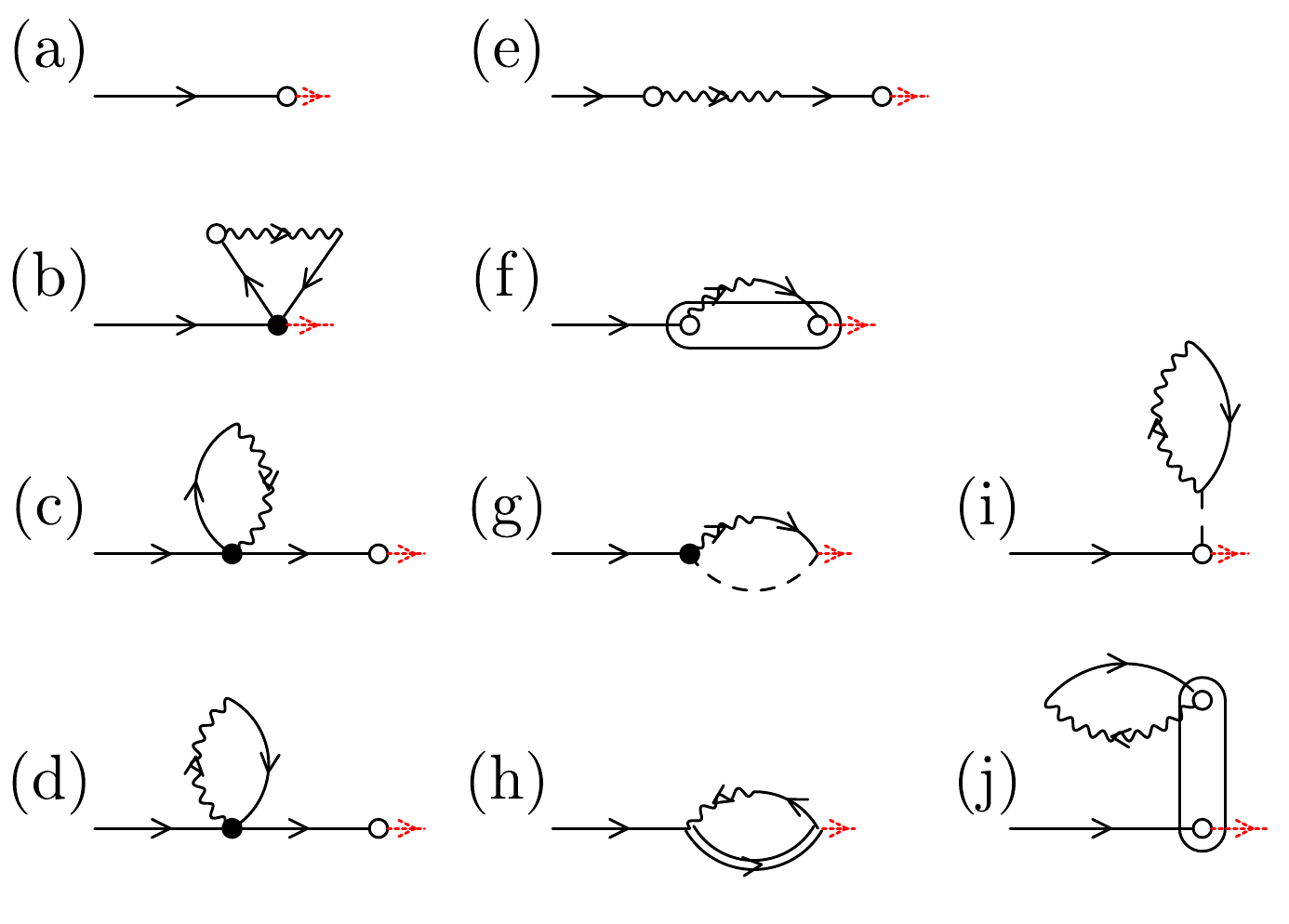}
\caption{\label{fig1}
Diagrams of the Hubbard NEGF (\ref{defG}) expansion in system-baths coupling.
Shown are (a) zero and (b-j) second order diagrams.
See text for details.
}
\end{figure}

Figure~\ref{fig1} shows diagrams of zero (panel a) and second (panels b-j)
order expansion in the system-bath coupling. 
Directed straight and wavy lines represent Fermion propagator $g_m(\tau,\tau')$
and self-energy $\sigma_{mm'}(\tau,\tau')$, Eq.~(\ref{sigma}),
respectively. Here $m\equiv S_1S_2$ is single electron transition between pair of many-body 
states $\lvert S_1\rangle$ and $\lvert S_2\rangle$, i.e. $N_{S_1}+1=N_{S_2}$
($N_S$ is number of electrons in state $\lvert S\rangle$).
Dashed line stands for Boson propagator $g_{b}(\tau,\tau')$ in the same charging block,
i.e. $b\equiv S_1S_2$ with $N_{S_1}=N_{S_2}$.
Directed double line represents two-electron propagator $d_b(\tau,\tau')$,
where $b\equiv S_1S_2$ with $N_{S_1}+2=N_{S_2}$.
Empty circle stands for (zero-order) spectral weight
\begin{equation}
 \langle \hat F_{m_1m_2}(\tau) \rangle_0 \equiv \big\langle \big\{\hat X_{m_1}(\tau);\hat X_{m_2}^\dagger(\tau)\big\}\big\rangle_0
\end{equation}
and oval with two circles is correlation function
\begin{equation}
\label{corr}
 \langle \delta \hat F_{m_1m_2}(\tau)\, \delta \hat F_{m_3m_4}(\tau')\rangle_0
\end{equation}
where $\delta \hat F_{mm'}=\hat F_{mm'}-\langle\hat F_{mm'}\rangle_0$.
Filled circles stand for `pruned' vertices~\cite{OvchinnikovValkov_2004}.
Finally, red dashed line indicates the end point of the diagram.
For more details see Ref.~\citenum{chen_nonequilibrium_2017}.

We note that Fig.~\ref{fig1} presents bare diagrams. As discussed in the previous
section, the Hubbard NEGF may be considered as expansion around results
of the Markov Redfield/Lindblad quantum master equation. Assuming we are dealing
with such an expansion, each diagram can be easily represented in terms of
state probabilities. For example, lesser and greater projections of the Fermion propagator 
$g_m(\tau,\tau')$ are ($m\equiv S_1S_2$ with $N_{S_1}+1=N_{S_2}$)
\begin{align}
 \label{gmlt_bare}
 g_m^{<}(t-t') &= i \frac{P_{S_2}}{P_{S_1}+P_{S_2}} e^{-i(E_{S_2}-E_{S_1})(t-t')}
 \\
 \label{gmgt_bare}
 g_m^{>}(t-t') &= -i\frac{P_{S_1}}{P_{S_1}+P_{S_2}} e^{-i(E_{S_2}-E_{S_1})(t-t')}
\end{align}
while its casual and anti-casual projections are 
\begin{align}
\label{gmc_bare}
 g_m^{c}(t-t') &= \theta(t-t') g_m^{>}(t-t') + \theta(t'-t) g_m^{<}(t-t')
 \\
 \label{gma_bare}
 g_m^{\tilde c}(t-t') &= \theta(t-t') g_m^{<}(t-t') + \theta(t'-t) g_m^{>}(t-t')
\end{align}
Similarly one can evaluate other elements of the diagrams, and represent the diagrams
in terms of states probabilities $P_S$ and time-dependent factors.  

We now turn to order-by-order analysis of the diagrams.
Before discussing the contributions we have to stress difference in Green function 
(Hilbert space) and quantum master equation (Liouville space) languages
mentioned also in our previous publication~\cite{galperin_comment_2015,galperin_photonics_2017}:
time arrangements on the Keldysh contour called diagrams in the QME language
are projections in the language of GFs. The difference is of minor importance for
the zero-order contribution, because the latter has only one diagram (see Fig.~\ref{fig1}a).
However, as is discussed later, it becomes critical for understanding higher order contributions 
and connection between the methods.

\begin{figure}[htbp]
\centering\includegraphics[width=\linewidth]{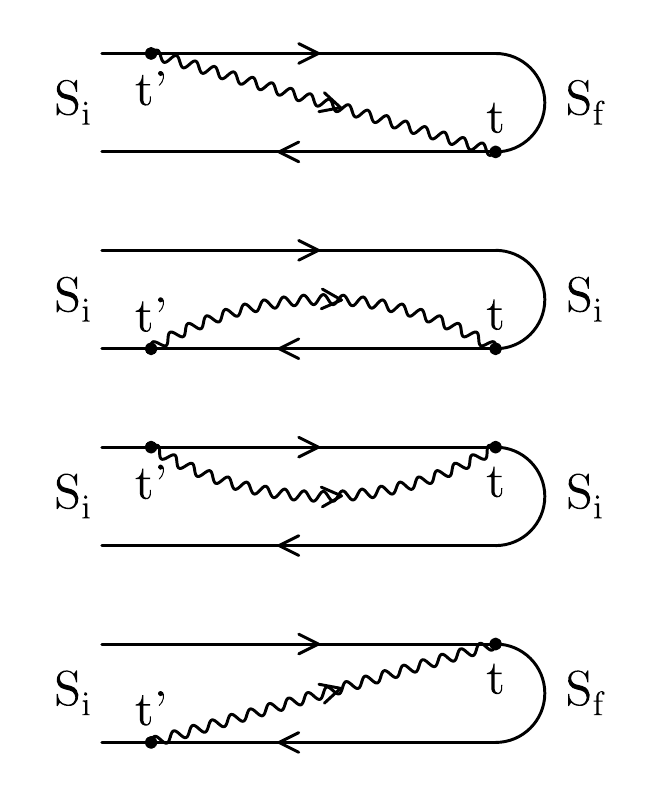}
\caption{\label{fig2}
Projections of the zero-order contribution, diagram in Fig.~\ref{fig1}a, to equation-of-motion 
(\ref{EOM}): the four projections correspond to the four terms in the right side of the 
expression.
}
\end{figure}

\subsection{Zero-order contributions}

Zero-order contribution to the Hubbard GF (\ref{defG}) is 
\begin{equation}
\label{ZERO}
 G_{mm'}^{(0)}(\tau,\tau')=g_m(\tau,\tau') \langle \hat F_{mm'}(\tau') \rangle_0
\end{equation}
Here $m=S_1S_2$ and $m'=S_3S_4$ with $N_{S_1}+1=N_{S_2}$ and $N_{S_3}+1=N_{S_4}$.
We note that there are four different projections of (\ref{ZERO}) contributing to
(\ref{EOM}) - one projection for each one of the terms on the right side of the expression.
The projections are given in Fig.~\ref{fig2}. 
Note there are eight projections when considering also complex conjugates 
taken into account by $2\,\mbox{Re}\ldots$ in (\ref{EOM}) - those are obtained by  interchanging time
positions between contour branches and flipping arrows.

Taking projections of Eq.~(\ref{ZERO}) on the Keldysh contour, substituting zero-order
expressions for the locator and spectral weight, Eqs.~(\ref{gmlt_bare})-(\ref{gma_bare}),
and utilizing the results in Eq.~(\ref{EOM}) yields expressions for the second order 
transfer rates from initial state $\lvert S_i\rangle$ to final state $\lvert S_f\rangle$
(see Supporting Information for details)
\begin{equation}
 W^{(2)}_{S_f\leftarrow S_i}=i\, \sigma^{>}_{S_fS_i,S_fS_i}(E_{S_i}-E_{S_f})
 -i\, \sigma^{<}_{S_iS_f,S_iS_f}(E_{S_f}-E_{S_i})
\end{equation}
As expected, these are golden rule type expressions.
We note in passing that while considering contributions in Eq.~(\ref{EOM}) from
different terms it is enough to account for rates appearing in front of $P_{S_i}$
with $S_i\neq S_f$, because rates of the type $S_i\leftarrow S_i$ (given by second and third
projections in Fig.~\ref{fig2}) can be obtained from the normalization condition
\begin{equation}
 \sum_{S_f} W_{S_f\leftarrow S_i} = 0
\end{equation}
The latter is easy to check also by direct evaluation of the rates.

\subsection{Second order contributions}
Diagrams of second order contributions to the Hubbard GF (\ref{defG}) are shown in 
panels (b)-(j) of Fig.~\ref{fig1}. Their explicit expressions are 
(letters next to expressions correspond to diagrams in Fig.~\ref{fig1})
\begin{align}
\label{GF2}
G_{mm'}^{(2)}(\tau,\tau') &= \sum_{m_1,m_2}\int_c d\tau_1\int_c d\tau_2\, 
\sigma_{m_1m_2}(\tau_1,\tau_2) \times
 \\
&\bigg( -i\, g_m(\tau,\tau')\, \gamma(m_2,mm')\, g_\gamma(\tau',\tau_1)\, 
\langle \hat F_{\gamma m_1}(\tau1)\rangle_0\, g_{m_2}(\tau_2,\tau'+)
& \mbox{(b)}
\nonumber \\
 &+i\, g_m(\tau,\tau_2)\, g_{m_1}(\tau_2,\tau_1)\, \gamma(\tilde m_1,m\tilde m_2)\,
 g_{\gamma}(\tau_2-,\tau')\,\langle \hat F_{\gamma m'}(\tau')\rangle_0
 & \mbox{(c)}
\nonumber \\
 & +i\, g_m(\tau,\tau_1)\, g_{m_2}(\tau_2,\tau_1+)\, \gamma(m_2,mm_1)\,
 g_\gamma(\tau_1,\tau')\,\langle\hat F_{\gamma m'}(\tau')\rangle_0
 & \mbox{(d)}
\nonumber \\
&+g_m(\tau,\tau_1)\,\langle\hat F_{mm_1}(\tau_1)\rangle_0\, g_{m_2}(\tau_2,\tau')\,
\langle\hat F_{m_2m'}(\tau')\rangle_0
& \mbox{(e)}
\nonumber \\
&+g_m(\tau,\tau_1)\, \langle T_c\,\delta\hat F_{mm_1}(\tau_1)\,\delta\hat F_{m_2m'}(\tau_2)\rangle_0\,
 g_{m_2}(\tau_2,\tau')
 & \mbox{(f)+(g)}
\nonumber \\
&+g_{m}(\tau,\tau_2)\,\langle T_c\,\delta\hat F_{m\tilde m_2}(\tau_2)\,\delta\hat F_{\tilde m_1m'}(\tau')\rangle_0\, g_{m_1}(\tau',\tau_1)
& \mbox{(h)}
\nonumber \\
 &- g_{m}(\tau,\tau')\, \langle T_c\,\delta\hat F_{mm'}(\tau')\,\delta\hat F_{m_2m_1}(\tau_1)\rangle_0\,
 g_{m_2}(\tau_2,\tau_1) 
 \bigg)
 & \mbox{(i)+(j)}
 \nonumber
\end{align}
Here $\tau-$ ($\tau+$) indicates contour variable right before (after) $\tau$ in contour ordering sense,
$\tilde m=S_2S_1$ for $m=S_1S_2$, and $\gamma$ is defined via
$\gamma(m_1,m_2m_3)\hat X_\gamma \equiv \left[\hat X_{m_1};\hat F_{m_2m_3}\right]$.

Second order brings into consideration two more contour variables, $\tau_1$ and $\tau_2$,
thus increasing number of projections. Moreover, Green function projections account only for
ordering of the variables on the Keldysh contour, while QME keeps track also on
ordering along real time axis. So, one Green function projection corresponds to
several QME projections~\cite{galperin_comment_2015,galperin_optical_2015}.
With four contour variables ($\tau$, $\tau'$, $\tau_1$, and $\tau_2$)  and with restriction
of $\tau$ (as variable representing the signal) being the latest time on real time axis
(causality principle) one has to account for $48$ QME projections.
Note, without $2\,\mbox{Re}\ldots$ term in (\ref{EOM}) number of QME projections doubles:
$3!$ orderings of times $t'$, $t_1$ and $t_2$ on the real time axes multiplied with
$2^4$ variants of distribution of the four variables on branches of the contour.
Each of terms in the right side of (\ref{EOM}) yields $12$ QME projections,
and each QME projection has contributions from several GF diagrams of Fig.~\ref{fig1}(b)-(j).

\begin{figure}[htbp]
\centering\includegraphics[width=\linewidth]{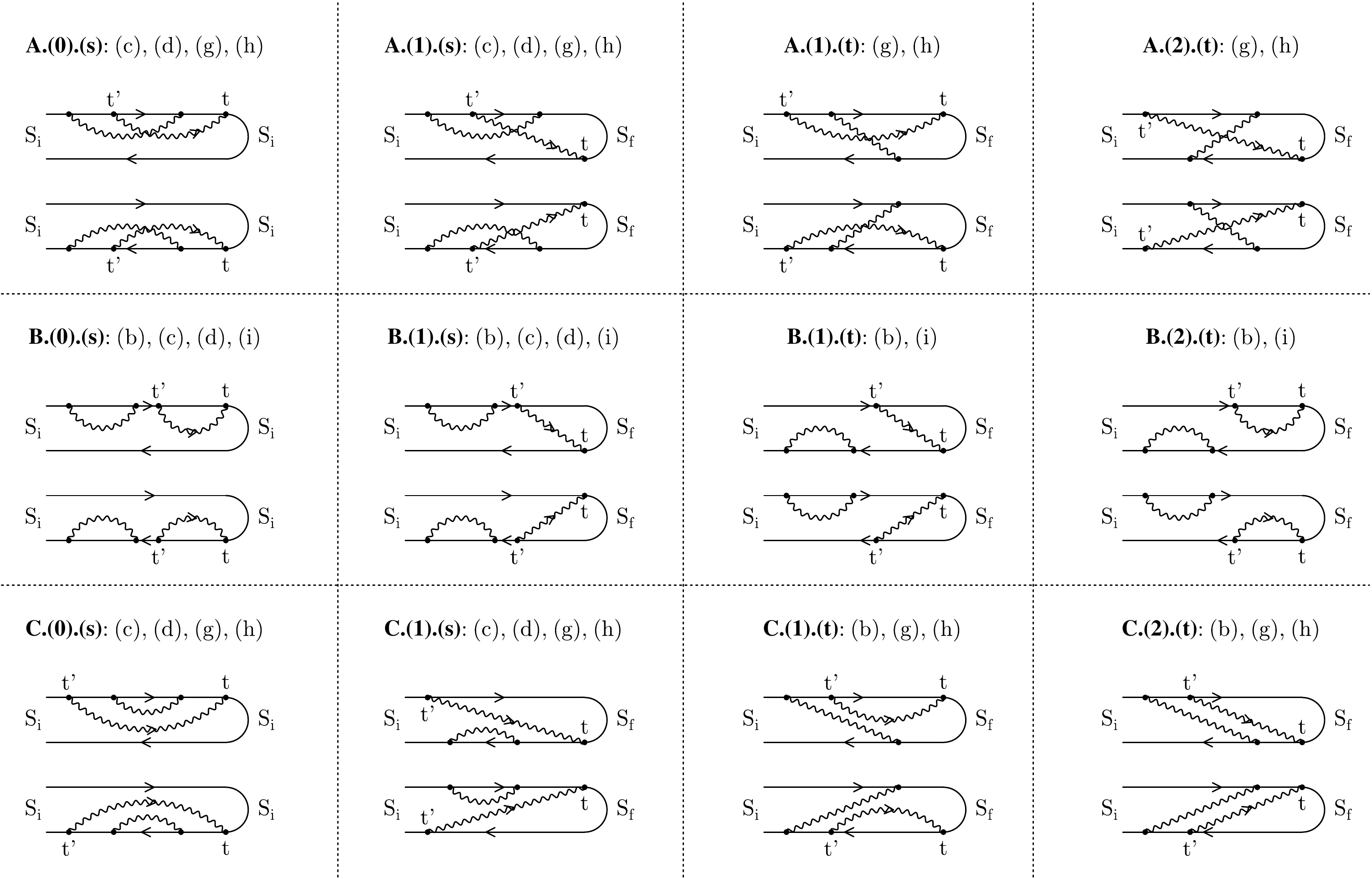}
\caption{\label{fig3}
Projections of the second-order contributions, diagrams in Figs.~\ref{fig1}(b)-(j), to equation-of-motion 
(\ref{EOM}). Following Ref.~\citenum{koller_density-operator_2010} 
the projections are classifies by 3 topologically different classes
($\bf A$, $\bf B$, $\bf C$), three groups ($\bf (0)$, $\bf (1)$, $\bf (2)$) 
and two sub-groups ($\bf (s)$ and $\bf (t)$).
Letters (b)-(j) correspond to second order diagrams from Fig.~\ref{fig1}
contributing to the projections.
See text for details.
}
\end{figure}

Figure~\ref{fig3} shows the projections distributed following 
Ref.~\citenum{koller_density-operator_2010} 
into three topologically different classes ($\bf A$, $\bf B$, $\bf C$) with three groups within each class
($\bf (0)$, $\bf (1)$ and $\bf (2)$ - minimum number of times on either branch of the contour)  
and with standalone ($\bf (s)$) or triple ($\bf (t)$) subgroups defined by number of different positions
latest time (except $t$, which is always the latest) can have relative to other times 
on the contour without changing projection topology. 
For simplicity, in each case we draw only one projection with latest time
being at the extreme right and did not indicate directions of self-energy line corresponding
to variables $\tau_1$ and $\tau_2$,
so that for the self-energy depending on directions of arrows
two different projections can be obtained from the one given in Fig.~\ref{fig3}.
As mentioned above, complex conjugate analogs of the considered diagrams
are obtained by inter-changing times between branches and flipping directions of lines.
We also indicate which of the diagrams from Fig.~\ref{fig1}(b)-(j) contribute to each of the
classes/groups/subgroups.
Substituting (\ref{GF2}) into (\ref{EOM}) yields expressions for the fourth order transfer rates.
Explicit expressions for the rates are given in the Supporting Information.

Comparing Hubbard NEGF approach to building transfer rates with the kinetic procedure
introduced in Refs.~\citenum{LeijnseWegewijsPRB08,koller_density-operator_2010} 
leads to several important observations as follows.
First, Hubbard NEGF diagrammatic technique naturally takes care of disregarding
disconnected diagrams, so that no additional special consideration is required.
Second, for dressed diagrams (see discussion in the next section)
artificial separation of the contributions into secular and non-secular parts
(depending on energy differences between pairs of states corresponding to transition 
relative to thermal energy) is not needed either.
Third, fourth order transfer rates built from bare expansion miss contributions from diagrams 
(e), (f), and (j) of Fig.~\ref{fig1}. The omission is natural because 
diagram (e) is responsible for deviations from steady-state (accumulation or depletion of
electrons in the system), which is absent at steady-state
considered in Refs.~\citenum{LeijnseWegewijsPRB08,koller_density-operator_2010}.
Diagrams (f) and (j) contain functions (\ref{corr}) which account for correlations 
between different states of the system. The latter are absent when rates are 
derived assuming independent (one-at-a-time) state population. 
Fourth, diagrams of Fig.~\ref{fig1} provide clear physical picture of transfer processes.
For example, diagram (g) is responsible for co-tunneling, while diagram (h) describes
two-electron transport. We stress that it is diagrams (not projections)
which are primary sources for the observed transfer characteristics.
Finally, combination of projections in order to simplify evaluation of rates
discussed in Ref.~\citenum{koller_density-operator_2010} is equivalent to change of 
consideration from QME-type projections, where real-time ordering is important,
to GF-type projections, where only ordering along the contour matters.
As indicated above, number of the latter projections is much smaller.
These observations indicate that GF based approach to building transfer rates is 
simpler and more efficient. In the next section we also argue that GFs also provide
an easy way to generalizations.

\subsection{Dressed rates}
Standard Green function diagrammatic procedure prescribes to dress the expansion diagrams
thus collecting a bare expansion into a resumed equation-of-motion. For the Hubbard NEGF
this leads to modified Dyson equation~\cite{OvchinnikovValkov_2004,chen_nonequilibrium_2017}.
However, full dressing makes transfer rates language obsolete. Thus, to go beyond 
bare expansion while still keeping kinetic scheme we propose to dress those diagram projections,
which were identified as contributing to rates in the bare diagrammatic expansion.
Technically, this means substituting zero-order expressions, Eqs.~(\ref{gmlt_bare})-(\ref{gma_bare}) 
for locators, spectral weights and correlation functions in the rates,
with corresponding dressed expressions (see Ref.~\citenum{chen_nonequilibrium_2017} 
and Supplementary Information for details). 
We note that while dressing one should be careful
to not introduce double counting. The dressing can be performed with respect 
to the same system-bath interaction in which original expansion was performed,
or with respect to interaction with other degrees of freedom/baths, or both.
Note that GF diagrammatic technique rules allow to account for additional interactions
in an ordered manner. This is contrary to {\em ad hoc} extensions considered in 
Refs.~\citenum{sumi_unified_2001,saito_unified_2009}. 


\section{Conclusion}\label{conclude}
We present Green function perspective on utilization of electron transfer techniques
in description of quantum transport. Central role in the former is played by transfer
rates, which are usually evaluated at the golden rule level of theory.
Necessity top generalize such rate expressions to account for intermediates was
indicated and {\em ad hoc} extension was formulated in Refs.~\citenum{sumi_unified_2001,saito_unified_2009}.
Similarly, kinetic schemes employing rates beyond second order in system-baths
coupling were discussed in Refs.~\citenum{LeijnseWegewijsPRB08,koller_density-operator_2010}.

We indicate that Green's function method is a natural way to account for
intermediates and go to any order in perturbative expansion in a well controlled way 
of diagrammatic expansion. However, inability to access system state resolved information
in the standard NEGF does not allow to establish connection between its diagrammatic 
expansion and rates utilized in kinetic schemes. At the same time, the Hubbard NEGF
theory, recently introduced by us~\cite{chen_nonequilibrium_2017}, yields such
a possibility. 

Utilizing zero-order (undressed) perturbation expansion of the Hubbard NEGF
up to second order in the system-bath coupling we establish connection with 
expressions of second and fourth order expressions for the rates introduced
in Refs.~\citenum{LeijnseWegewijsPRB08,koller_density-operator_2010}.
We discuss connection between the two approaches and indicate advantages of Green's 
function techniques in deriving expressions for the rates. Finally, we note that
standard dressing of diagrams in Green's function expansions yields a possibility
of formulating generalized (dressed) expressions for rates capable to account for
additional baths (degrees of freedom) in a well controlled ordered manner.
Practical application of the theory to realistic simulations is a goal for future research.



\begin{acknowledgement}
This material is based upon work supported by the National Science Foundation 
under CHE-1565939.
\end{acknowledgement}

\begin{suppinfo}
The Supporting Information is available free of charge on the ACS Publications website at DOI: XXX

Explicit expressions for second and fourth order transfer rates.
\end{suppinfo}


\providecommand{\latin}[1]{#1}
\makeatletter
\providecommand{\doi}
  {\begingroup\let\do\@makeother\dospecials
  \catcode`\{=1 \catcode`\}=2\doi@aux}
\providecommand{\doi@aux}[1]{\endgroup\texttt{#1}}
\makeatother
\providecommand*\mcitethebibliography{\thebibliography}
\csname @ifundefined\endcsname{endmcitethebibliography}
  {\let\endmcitethebibliography\endthebibliography}{}

\begin{tocentry}
\centering\includegraphics{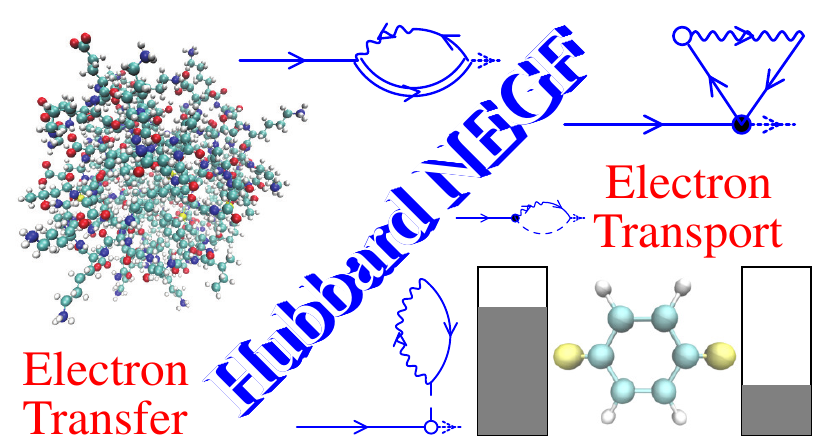}
\end{tocentry}

\end{document}


\maketitle

Here we give explicit expressions for second and fourth order transfer rates: 
first in terms of locators, spectral weights and correlation functions;
after that, result of substitution of zero order expressions for the latter.
results of Refs.~\citenum{LeijnseWegewijsPRB08,koller_density-operator_2010} for $W^{(n)}_{S_f\leftarrow S_i}$ rate is obtained
by setting  $P_{S_i}=1$ and all other probabilities to $0$.

\section{Second Order Rates}
The only diagram contributing to second order rate is shown in Fig.~1a of the main text.

First and fourth terms in the right side of Eq.~(11) in the main text
\begin{align*}
& 2\,\mbox{Re}\sum_{K\in\{L,R\}}\int_{-\infty}^t dt'\,
\\ &\bigg(\ 
\sigma^{K\, >}_{S_fS_i,S_fS_i}(t-t')\, g^{<}_{S_fS_i}(t'-t)\,\langle\hat F_{S_fS_i,S_fS_i}(t)\rangle
\\
&+
\sigma^{K\, <}_{S_iS_f,S_iS_f}(t-t')\, g^{>}_{S_iS_f}(t'-t)\,\langle\hat F_{S_iS_f,S_iS_f}(t)\rangle
\bigg)
\end{align*}
contribute to $W^{(2)}_{S_f\leftarrow S_i}\, P_{S_i}$.
Explicit expression for the rate is 
\begin{equation*}
 W^{(2)}_{S_f\leftarrow S_i} = i\sum_{K\in\{L,R\}}
 \bigg(
 \sigma^{K\, >}_{S_fS_i,S_fS_i}(E_{S_i}-E_{S_f}) - \sigma^{K\, <}_{S_iS_f,S_iS_f}(E_{S_f}-E_{S_i})
 \bigg)
\end{equation*}

Second and third terms in the right side of Eq.~(11) in the main text
\begin{align*}
& -2\,\mbox{Re}\sum_{S}\sum_{K\in\{L,R\}}\int_{-\infty}^t dt'\,
\\ &\bigg(\ 
\sigma^{K\, >}_{SS_i,SS_i}(t-t')\, g^{<}_{SS_i}(t'-t)\,\langle\hat F_{SS_i,SS_i}(t)\rangle
\\
&+
\sigma^{K\, <}_{S_iS,S_iS}(t-t')\, g^{>}_{S_iS}(t'-t)\,\langle\hat F_{S_iS,S_iS}(t)\rangle
\bigg)
\end{align*}
contribute to $W^{(2)}_{S_i\leftarrow S_i}\, P_{S_i}$.
Explicit expression for the rate is 
\begin{equation*}
 W^{(2)}_{S_i\leftarrow S_i} = -i\sum_{S}\sum_{K\in\{L,R\}}
 \bigg(
 \sigma^{K\, >}_{SS_i,SS_i}(E_{S_i}-E_{S}) - \sigma^{K\, <}_{S_iS,S_iS}(E_{S}-E_{S_i})
 \bigg)
\end{equation*}

It is easy to check that 
\[
 \sum_{S_f} W^{(2)}_{S_f\leftarrow S_i} = 0
\]

\section{Fourth Order Rates}
Contributions are classified by projection in Fig.~3 and diagram in Fig.~1
of the main text. 
In the expressions below
\begin{align*}
 B_{SS'}(\tau,\tau') &\equiv -i\langle \hat X_{SS'}(\tau)\, \hat X^\dagger_{SS'}(\tau')\rangle
 & N_{S}=N_{S'}
 \\
 D_{SS'}(\tau,\tau') &\equiv -i\langle \hat X_{SS'}(\tau)\,\hat X_{SS'}^\dagger(\tau')\rangle
 & N_{S}+2=N_{S'}
\end{align*}

\subsection{A.(0).(s) projections}
This contribution is of the type
$
 W^{(4)}_{S_i\leftarrow S_i}\, P_{S_i}
$
\subsubsection{Diagram (c)}
\begin{align*}
&2\,\mbox{Im}\sum_{S_1,S_2,S_3}\sum_{K_1,K_2\in\{L,R\}}
\int_{-\infty}^{t}dt_1\int_{-\infty}^{t_1}dt'\int_{-\infty}^{t'}dt_2\,
\\
&\bigg(\ 
\sigma^{K_1\, <}_{S_iS_1,S_3S_2}(t-t')\, g_{S_3S_2}^{>}(t'-t_1)\, g_{S_3S_i}^{<}(t_1-t_2)\,
\sigma^{K_2\, >}_{S_3S_i,S_2S_1}(t_2-t_1)\, g_{S_iS_1}^{>}(t_1-t)
\langle \hat F_{S_iS_1,S_iS_1}(t)\rangle
\\
&-
\sigma^{K_1\, >}_{S_1S_i,S_2S_3}(t-t')\, g^{<}_{S_2S_3}(t'-t_1)\, g^{>}_{S_iS_3}(t_1-t_2)\,
\sigma^{K_2\, <}_{S_iS_3,S_1S_2}(t_2-t_1)\, g^{<}_{S_1S_i}(t_1-t)\langle\hat F_{S_1S_i,S_1S_i}(t)\rangle
\\
&+
\sigma^{K_1\, >}_{S_1S_i,S_2S_3}(t-t')\, g^{>}_{S_2S_3}(t'-t_2)\,g^{<}_{S_2S_1}(t_2-t_1)\,
 \sigma^{K_2\, >}_{S_2S_1,S_3S_i}(t_1-t_2)\, g^{<}_{S_1S_i}(t_2-t)\langle \hat F_{S_1S_i,S_1S_i}(t)\rangle
\\
&-
\sigma^{K_1\, <}_{S_iS_1,S_3S_2}(t-t')\, g^{<}_{S_3S_2}(t'-t_2)\, g^{>}_{S_1S_2}(t_2-t_1)\,
\sigma^{K_2\, <}_{S_1S_2,S_iS_3}(t_1-t_2)\, g^{>}_{S_iS_1}(t_2-t)\langle\hat F_{S_iS_1,S_iS_1}(t)\rangle
\bigg)
\end{align*}

\subsubsection{Diagram (d)}
\begin{align*}
&2\, \mbox{Im}\sum_{S_1,S_2,S_3}\sum_{K_1,K_2\in\{L,R\}}
\int_{-\infty}^t dt_1\int_{-\infty}^{t_1} dt'\int_{-\infty}^{t'}dt_2\,
\\
&\bigg(\ 
\sigma^{K_1\, >}_{S_1S_i,S_2S_3}(t-t')\, g^{>}_{S_2S_3}(t'-t_2)\,\sigma^{K_2\, <}_{S_iS_3,S_1S_2}(t_2-t_1)\,
 g^{>}_{S_1S_2}(t_1-t_2)\, g^{<}_{S_1S_i}(t_2-t)\langle\hat F
_{S_1S_i,S_1S_i}(t)\rangle
\\
&-
\sigma^{K_1\, <}_{S_iS_1,S_3S_2}(t-t')\, g_{S_3S_2}^{<}(t'-t_2)\,\sigma^{K_2\, >}_{S_3S_i,S_2S_1}(t_2-t_1)\,
 g^{<}_{S_2S_1}(t_1-t_2)\, g^{>}_{S_iS_1}(t_2-t)\,\langle\hat F_{S_iS_1,S_iS_1}(t)\rangle
 \\
&+
\sigma^{K_1\, <}_{S_iS_1,S_3S_2}(t-t')\, g^{>}_{S_3S_2}(t'-t_1)\,\sigma^{K_2\, <}_{S_1S_2,S_iS_3}(t_1-t_2)\,
 g^{>}_{S_iS_3}(t_2-t_1)\, g^{>}_{S_iS_1}(t_1-t)\langle\hat F_{S_iS_1,S_iS_1}(t)\rangle
 \\
&-
\sigma^{K_1\, >}_{S_1S_i,S_2S_3}(t-t')\, g^{<}_{S_2S_3}(t'-t_1)\,\sigma^{K_2\, >}_{S_2S_1,S_3S_i}(t_1-t_2)\,
 g^{<}_{S_3S_i}(t_2-t_1)\, g^{<}_{S_1S_i}(t_1-t)\langle\hat F_{S_1S_i,S_1S_i}(t)\rangle
\bigg)
\end{align*}

\subsubsection{Diagram (g)}
\begin{align*}
&2\,\mbox{Im}\sum_{S_1,S_2,S_3}\sum_{K_1,K_2\in\{L,R\}}\int_{-\infty}^t dt_1\int_{-\infty}^{t_1} dt'\int_{-\infty}^{t'}dt_2
\\
&\bigg(
\sigma^{K_1\, <}_{S_iS_1,S_3S_2}(t-t')\, g^{<}_{S_3S_2}(t'-t_2)\,\sigma^{K_2\, >}_{S_3S_i,S_2S_1}(t_2-t_1)\, g^{>}_{S_2S_1}(t_1-t)\, B^{<}_{S_2S_i}(t-t_2)
\\
&+
\sigma^{K_1\, >}_{S_1S_i,S_2S_3}(t-t')\, g^{>}_{S_2S_3}(t'-t_2)\,\sigma^{K_2\, <}_{S_iS_3,S_1S_2}(t_2-t_1)\, g^{<}_{S_1S_2}(t_1-t)\, B^{>}_{S_iS_2}(t-t_2)
\bigg)
\end{align*}

\subsubsection{Diagram (h)}
\begin{align*}
&2\,\mbox{Im}\sum_{S_1,S_2,S_3}\sum_{K_1,K_2\in\{L,R\}}\int_{-\infty}^tdt_1\int_{-\infty}^{t_1} dt'\int_{-\infty}^{t'}dt_2
\\
&\bigg(
\sigma^{K_1\, <}_{S_iS_1,S_3S_2}(t-t')\, g^{<}_{S_3S_2}(t'-t_2)\, D_{S_iS_2}^{>}(t_2-t)\, g^{<}_{S_1S_2}(t-t_1)\,\sigma^{K_2\, <}_{S_1S_2,S_iS_3}(t_1-t_2)
\\
&+
\sigma^{K_1\, >}_{S_1S_i,S_2S_3}(t-t')\, g^{>}_{S_2S_3}(t'-t_2)\, D^{<}_{S_2S_i}(t_2-t)\, g^{>}_{S_2S_1}(t-t_1)\,\sigma^{K_2\, >}_{S_2S_1,S_3S_i}(t_1-t_2)\bigg)
\end{align*}

\subsubsection{Resulting zero-order expression}
\begin{align*}
&-2\,\mbox{Im} \sum_{S_1,S_2,S_3}\sum_{K_1,K_2\in\{L,R\}}\int\frac{d\omega_1}{2\pi}\int\frac{d\omega_2}{2\pi}\,
 \bigg(\frac{P_{S_i}\, P_{S_3}}{P_{S_i}+P_{S_3}} + P_{S_2}\bigg) \frac{P_{S_i}}{P_{S_2}+P_{S_3}}
\\
&\bigg[\ 
\frac{
\Gamma^{K_1}_{S_iS_1,S_3S_2}(\omega_1)f_{K_1}(\omega_1)
\Gamma^{K_2}_{S_3S_i,S_2S_1}(\omega_2)[1-f_{K_2}(\omega_2)]}
{(E_{S_3}-E_{S_i}+\omega_2+i0)(E_{S_2}-E_{S_i}-\omega_1+\omega_2+i0)(E_{S_1}-E_{S_i}-\omega_1+i0)}
\\
&+\frac{
\Gamma^{K_1}_{S_iS_1,S_3S_2}(\omega_1)f_{K_1}(\omega_1)
\Gamma^{K_2}_{S_1S_2,S_iS_3}(\omega_2)f_{K_2}(\omega_2)}
{(E_{S_3}-E_{S_i}-\omega_2+i0)(E_{S_2}-E_{S_i}-\omega_1-\omega_2+i0)(E_{S_1}-E_{S_i}-\omega_1+i0)}
\\
&+\frac{
\Gamma^{K_1}_{S_1S_i,S_2S_3}(\omega_1)[1-f_{K_1}(\omega_1)]
\Gamma^{K_2}_{S_2S_1,S_3S_i}(\omega_2)[1-f_{K_2}(\omega_2)]}
{(E_{S_i}-E_{S_3}-\omega_2+i0)(E_{S_i}-E_{S_2}-\omega_1-\omega_2+i0)(E_{S_i}-E_{S_1}-\omega_1+i0)}
\\
&+\frac{
\Gamma^{K_1}_{S_1S_i,S_2S_3}(\omega_1)[1-f_{K_1}(\omega_1)]
\Gamma^{K_2}_{S_iS_3,S_1S_2}(\omega_2)f_{K_2}(\omega_2)}
{(E_{S_i}-E_{S_3}+\omega_2+i0)(E_{S_i}-E_{S_2}-\omega_1+\omega_2+i0)(E_{S_i}-E_{S_1}-\omega_1+i0)}
\bigg]
\end{align*}


\subsection{A.(1).(s) projections}
This contribution is of the type
$
 W^{(4)}_{S_f\leftarrow S_i}\, P_{S_i}
$

\subsubsection{Diagram (c)}
\begin{align*}
&-2\,\mbox{Im}\sum_{S_1,S_2}\sum_{K_1,K_2\in\{L,R\}}
\int_{-\infty}^t dt_1\int_{-\infty}^{t_1} dt' \int_{-\infty}^{t'} dt_2\,
\\
&\bigg(\ 
\sigma^{K_1\, >}_{S_fS_i,S_2S_1}(t-t')\, g^{>}_{S_2S_1}(t'-t_2)\, g^{<}_{S_2S_f}(t_2-t_1)\,
\sigma^{K_2\, >}_{S_2S_f,S_1S_i}(t_1-t_2)\, g^{<}_{S_fS_i}(t_2-t)\langle\hat F_{S_fS_i,S_fS_i}(t)\rangle
\\
&-
\sigma^{K_1\, <}_{S_iS_f,S_1S_2}(t-t')\, g^{<}_{S_1S_2}(t'-t_2)\, g^{>}_{S_fS_2}(t_2-t_1)\,
\sigma^{K_2\, <}_{S_fS_2,S_iS_1}(t_1-t_2)\, g^{>}_{S_iS_f}(t_2-t)\langle\hat F_{S_iS_f,S_iS_f}(t)\rangle
\\
&+
\sigma^{K_1\, <}_{S_iS_f,S_1S_2}(t-t')\, g^{>}_{S_1S_2}(t'-t_1)\, g^{<}_{S_1S_i}(t_1-t_2)\,
\sigma^{K_2\, >}_{S_1S_i,S_2S_f}(t_2-t_1)\, g^{>}_{S_iS_f}(t_1-t)\langle\hat F_{S_iS_f,S_iS_f}(t)\rangle
\\
&-
\sigma^{K_1\, >}_{S_fS_i,S_2S_1}(t-t')\, g^{<}_{S_2S_1}(t'-t_1)\, g^{>}_{S_iS_1}(t_1-t_2)\,
\sigma^{K_2\, <}_{S_iS_1,S_fS_2}(t_2-t_1)\, g^{<}_{S_fS_i}(t_1-t)\langle\hat F_{S_fS_i,S_fS_i}(t)\rangle
\bigg)
\end{align*}

\subsubsection{Diagram (d)}
\begin{align*}
&-2\,\mbox{Im}\sum_{S_1,S_2}\sum_{K_1,K_2\in\{L,R\}}
\int_{-\infty}^t dt_1\int_{-\infty}^{t_1}dt'\int_{-\infty}^{t'} dt_2\,
\\
&\bigg(\ 
\sigma^{K_1\, <}_{S_iS_f,S_1S_2}(t-t')\, g^{>}_{S_1S_2}(t'-t_1)\, 
\sigma^{K_2\, <}_{S_fS_2,S_iS_1}(t_1-t_2)\,  
g^{>}_{S_iS_1}(t_2-t_1)\, g^{>}_{S_iS_f}(t_1-t)\langle\hat F_{S_iS_f,S_iS_f}(t)\rangle
\\
&-
\sigma^{K_1\, >}_{S_fS_i,S_2S_1}(t-t')\, g^{<}_{S_2S_1}(t'-t_1)\, 
\sigma^{K_2\, >}_{S_2S_f,S_1S_i}(t_1-t_2)\, 
g^{<}_{S_1S_i}(t_2-t_1)\, g^{<}_{S_fS_i}(t_1-t)\langle\hat F_{S_fS_i,S_f,S_i}(t)\rangle
\\
 &+
\sigma^{K_1\, >}_{S_fS_i,S_2S_1}(t-t')\, g^{>}_{S_2S_1}(t'-t_2)\, 
 \sigma^{K_2\, <}_{S_iS_1,S_fS_2}(t_2-t_1)\, 
 g^{>}_{S_fS_2}(t_1-t_2)\, g^{<}_{S_fS_i}(t_2-t)\langle\hat F_{S_fS_i,S_fS_i}(t)\rangle
  \\
&-
\sigma^{K_1\, <}_{S_iS_f,S_1S_2}(t-t')\, g^{<}_{S_1S_2}(t'-t_2)\, 
\sigma^{K_2\, >}_{S_1S_i,S_2S_f}(t_2-t_1)\, 
g^{<}_{S_2S_f}(t_1-t_2)\, g^{>}_{S_iS_f}(t_2-t)\langle\hat F_{S_iS_f,S_iS_f}(t)\rangle
 \bigg)
\end{align*}

\subsubsection{Diagram (g)}
\begin{align*}
&-2\,\mbox{Im}\sum_{S_1,S_2}\sum_{K_1,K_2\in\{L,R\}}
\int_{-\infty}^t dt_1\int_{-\infty}^{t_1} dt'\int_{-\infty}^{t'} dt_2
\\
&\bigg(\ 
\sigma^{K_1\, >}_{S_fS_i,S_2S_1}(t-t')\, g^{>}_{S_2S_1}(t'-t_2)\, 
 \sigma^{K_2\, <}_{S_iS_1,S_fS_2}(t_2-t_1)\, g^{<}_{S_fS_2}(t_1-t)\,
 B^{>}_{S_iS_2}(t-t_2)
\\
&+
\sigma^{K_1\, <}_{S_iS_f,S_1S_2}(t-t')\, g^{<}_{S_1S_2}(t'-t_2)\, 
\sigma^{K_2\, >}_{S_1S_i,S_2S_f}(t_2-t_1)\, g^{>}_{S_2S_f}(t_1-t)\,
B^{<}_{S_2S_i}(t-t_2)
\bigg)
 \end{align*}

\subsubsection{Diagram (h)}
\begin{align*}
&-2\,\mbox{Im}\sum_{S_1,S_2}\sum_{K_1,K_2\in\{L,R\}}
\int_{-\infty}^t dt_1\int_{-\infty}^{t_1} dt' \int_{-\infty}^{t'} dt_2
\\
&\bigg(
\sigma^{K_1\, >}_{S_fS_i,S_2S_1}(t-t')\, g^{>}_{S_2S_1}(t'-t_2)\, D^{<}_{S_2S_i}(t_2-t)\,
g^{>}_{S_2S_f}(t-t_1)\,\sigma^{K_2\, >}_{S_2S_f,S_1S_i}(t_1-t_2)
\\
& +
 \sigma^{K_1\, <}_{S_iS_f,S_1S_2}(t-t')\, g^{<}_{S_1S_2}(t'-t_2)\, 
D^{>}_{S_iS_2}(t_2-t)\, g^{<}_{S_fS_2}(t-t_1)\, \sigma^{K_2\, <}_{S_fS_2,S_iS_1}(t_1-t_2)
\bigg)
\end{align*}

\subsubsection{Resulting zero-order expression}
\begin{align*}
&2\,\mbox{Im}\sum_{S_1,S_2}\sum_{K_1,K_2\in\{L,R\}}
\int\frac{d\omega_1}{2\pi}\int\frac{d\omega_2}{2\pi}\,
\bigg(P_{S_2}+\frac{P_{S_i}P_{S_1}}{P_{S_i}+P_{S_1}}\bigg) \frac{P_{S_i}}{P_{S_1}+P_{S_2}}
\\
&\bigg[\ 
\frac{
\Gamma^{K_1}_{S_fS_i,S_2S_1}(\omega_1)[1-f_{K_1}(\omega_1)]
\Gamma^{K_2}_{S_2S_f,S_1S_i}(\omega_2)[1-f_{K_2}(\omega_2)]}
{(E_{S_i}-E_{S_1}-\omega_2+i0)(E_{S_i}-E_{S_2}-\omega_1-\omega_2+i0)(E_{S_i}-E_{S_f}-\omega_1+i0)}
\\
&+
\frac{
\Gamma^{K_1}_{S_fS_i,S_2S_1}(\omega_1)[1-f_{K_1}(\omega_1)]
\Gamma^{K_2}_{S_iS_1,S_fS_2}(\omega_2)f_{K_2}(\omega_2)}
{(E_{S_i}-E_{S_1}+\omega_2+i0)(E_{S_i}-E_{S_2}-\omega_1+\omega_2+i0)(E_{S_i}-E_{S_f}-\omega_1+i0)}
\\
&+
\frac{
\Gamma^{K_1}_{S_iS_f,S_1S_2}(\omega_1)f_{K_1}(\omega_1)
\Gamma^{K_2}_{S_1S_i,S_2S_f}(\omega_2)[1-f_{K_2}(\omega_2)]}
{(E_{S_1}-E_{S_i}+\omega_2+i0)(E_{S_2}-E_{S_i}-\omega_1+\omega_2+i0)(E_{S_f}-E_{S_i}-\omega_1+i0)}
\\
&+
\frac{
\Gamma^{K_1}_{S_iS_f,S_1S_2}(\omega_1)f_{K_1}(\omega_1)
\Gamma^{K_2}_{S_fS_2,S_iS_1}(\omega_2)f_{K_2}(\omega_2)}
{(E_{S_1}-E_{S_i}-\omega_2+i0)(E_{S_2}-E_{S_i}-\omega_1-\omega_2+i0)(E_{S_f}-E_{S_i}-\omega_1+i0)}
\bigg]
\end{align*}


\subsection{A.(1).(t) projections}
This contribution is of the type
$
 W^{(4)}_{S_f\leftarrow S_i}\, P_{S_i}
$

\subsubsection{Diagram (g)}
\begin{align*}
& -2\,\mbox{Im}\sum_{S_1,S_2}\sum_{K_1,K_2\in\{L,R\}}
\int_{-\infty}^{t}dt_1\int_{-\infty}^{t_1}dt'\int_{-\infty}^t dt_2
\\
&\bigg(\ 
\sigma^{K_1\, <}_{S_fS_2,S_iS_1}(t-t')\, g^{>}_{S_iS_1}(t'-t_1)\,\sigma^{K_2\, >}_{S_2S_1,S_fS_i}(t_1-t_2)\,
g^{<}_{S_fS_i}(t_2-t)\, B^{<}_{S_2S_i}(t-t_1)
\\
& +
\sigma^{K_1\, >}_{S_2S_f,S_iS_1}(t-t')\, g^{<}_{S_1S_i}(t'-t_1)\,\sigma^{K_2\, <}_{S_1S_2,S_iS_f}(t_1-t_2)\,
g^{>}_{S_iS_f}(t_2-t)\, B^{>}_{S_iS_2}(t-t_1)
\bigg)
\end{align*}

\subsubsection{Diagram (h)}
\begin{align*}
& - 2\,\mbox{Im}\sum_{S_1,S_2}\sum_{K_1,K_2\in\{L,R\}} 
\int_{-\infty}^{t} dt_1 \int_{-\infty}^{t_1} dt' \int_{-\infty}^t dt_2
\\
& \bigg(\ 
\sigma^{K_1\, <}_{S_fS_2,S_iS_1}(t-t')\, g^{>}_{S_iS_1}(t'-t_1)\, D^{>}_{S_iS_2}(t_1-t)\,
g^{>}_{S_iS_f}(t-t_2)\,\sigma^{K_2\, <}_{S_iS_f,S_1S_2}(t_2-t_1)
\\
& +
\sigma^{K_1\, >}_{S_2S_f,S_1S_i}(t-t')\, g^{<}_{S_1S_i}(t'-t_1)\, D^{<}_{S_2S_i}(t_1-t)\,
g^{<}_{S_fS_i}(t-t_2)\,\sigma^{K_2\, >}_{S_fS_i,S_2S_1}(t_2-t_1)
\bigg)
\end{align*}

\subsubsection{Resulting zero-order expression}
\begin{align*}
&2\,\mbox{Im}\sum_{S_1,S_2}\sum_{K_1,K_2\in\{L,R\}}
\int\frac{d\omega_1}{2\pi}\int\frac{d\omega_2}{2\pi} \,
\frac{P_{S_i}^3}{(P_{S_i}+P_{S_f})(P_{S_i}+P_{S_1})}
\\
&\bigg[\ 
\frac{
\Gamma^{K_1}_{S_fS_2,S_iS_1}(\omega_1)f_{K_1}(\omega_1)
\Gamma^{K_2}_{S_2S_1,S_fS_i}(\omega_2)[1-f_{K_2}(\omega_2)]}
{(E_{S_1}-E_{S_i}-\omega_1+i0)(E_{S_2}-E_{S_i}-\omega_1+\omega_2+i0)(E_{S_i}-E_{S_f}-\omega_2+i0)}
\\
& +
\frac{
\Gamma^{K_1}_{S_fS_2,S_iS_1}(\omega_1)f_{K_1}(\omega_1)
\Gamma^{K_2}_{S_iS_f,S_1S_2}(\omega_2)f_{K_2}(\omega_2)}
{(E_{S_1}-E_{S_i}-\omega_1+i0)(E_{S_2}-E_{S_i}-\omega_1-\omega_2+i0)(E_{S_i}-E_{S_f}+\omega_2+i0)}
\\
& +
\frac{
\Gamma^{K_1}_{S_2S_f,S_1S_i}(\omega_1)[1-f_{K_1}(\omega_1)]
\Gamma^{K_2}_{S_fS_i,S_2S_1}(\omega_2)[1-f_{K_2}(\omega_2)]}
{(E_{S_i}-E_{S_1}-\omega_1+i0)(E_{S_i}-E_{S_2}-\omega_1-\omega_2+i0)(E_{S_f}-E_{S_i}+\omega_2+i0)}
\\
& +
\frac{
\Gamma^{K_1}_{S_2S_f,S_iS_1}(\omega_1)[1-f_{K_1}(\omega_1)]
\Gamma^{K_2}_{S_1S_2,S_iS_f}(\omega_2)f_{K_2}(\omega_2)}
{(E_{S_i}-E_{S_1}-\omega_1+i0)(E_{S_i}-E_{S_2}-\omega_1+\omega_2+i0)(E_{S_f}-E_{S_i}-\omega_2+i0)}
\bigg]
\end{align*}


\subsection{A.(2).(t) projections}
This contribution is of the type
$
 W^{(4)}_{S_f\leftarrow S_i}\, P_{S_i}
$

\subsubsection{Diagram (g)}
\begin{align*}
&2\,\mbox{Im}\sum_{S_1,S_2}\sum_{K_1,K_2\in\{L,R\}}
\int_{-\infty}^t dt_1\int_{-\infty}^{t_1} dt' \int_{-\infty}^t dt_2
\\
&\bigg(\ 
\sigma^{K_1\, >}_{S_fS_2,S_1S_i}(t-t')\, g^{<}_{S_1S_i}(t'-t_1)\, \sigma^{K_2\, <}_{S_1S_f,S_iS_2}(t_1-t_2)\,
g^{>}_{S_iS_2}(t_2-t)\, B^{>}_{S_iS_f}(t-t_1)
\\
&+
\sigma^{K_1\, <}_{S_2S_f,S_iS_1}(t-t')\, g^{>}_{S_iS_1}(t'-t_1)\, \sigma^{K_2\, >}_{S_fS_1,S_2S_i}(t_1-t_2)\,
g^{<}_{S_2S_i}(t_2-t)\, B^{<}_{S_fS_i}(t-t_1)
\bigg)
\end{align*}

\subsubsection{Diagram (h)}
\begin{align*}
&2\,\mbox{Im}\sum_{S_1,S_2}\sum_{K_1,K_2\in\{L,R\}}
\int_{-\infty}^t dt_1 \int_{-\infty}^{t_1} dt' \int_{-\infty}^t dt_2
\\
&\bigg(\ 
\sigma^{K_1\, >}_{S_fS_2,S_1S_i}(t-t')\, g^{<}_{S_1S_i}(t'-t_1)\, D^{<}_{S_fS_i}(t_1-t)\,
g^{<}_{S_2S_i}(t-t_2)\,\sigma^{K_2\, >}_{S_2S_i,S_fS_1}(t_2-t_1)
\\
&+
\sigma^{K_1\, <}_{S_2S_f,S_iS_1}(t-t')\, g^{>}_{S_iS_1}(t'-t_1)\, D^{>}_{S_iS_f}(t_1-t)\,
g^{>}_{S_iS_2}(t-t_2)\,\sigma^{K_2\, <}_{S_iS_2,S_1S_f}(t_2-t_1)
\bigg)
\end{align*}

\subsubsection{Resulting zero-order expression}
\begin{align*}
&-2\,\mbox{Im}\sum_{S_1,S_2}\sum_{K_1,K_2\in\{L,R\}} 
\int\frac{d\omega_1}{2\pi}\int \frac{d\omega_2}{2\pi}\,
\frac{P_{S_i}^3}{(P_{S_i}+P_{S_1})(P_{S_i}+P_{S_2})}
\\
&\bigg[\ 
\frac{
\Gamma^{K_1}_{S_fS_2,S_1S_i}(\omega_1)[1-f_{K_1}(\omega_1)]
\Gamma^{K_2}_{S_2S_i,S_fS_1}(\omega_2)[1-f_{K_2}(\omega_2)]}
{(E_{S_i}-E_{S_1}-\omega_1+i0)(E_{S_i}-E_{S_f}-\omega_1-\omega_2+i0)(E_{S_2}-E_{S_i}+\omega_2+i0)}
\\
&+
\frac{
\Gamma^{K_1}_{S_fS_2,S_1S_i}(\omega_1)[1-f_{K_1}(\omega_1)]
\Gamma^{K_2}_{S_1S_f,S_iS_2}(\omega_2)f_{K_2}(\omega_2)}
{(E_{S_i}-E_{S_1}-\omega_1+i0)(E_{S_i}-E_{S_f}-\omega_1+\omega_2+i0)(E_{S_2}-E_{S_i}-\omega_2+i0)}
\\
&+
\frac{
\Gamma^{K_1}_{S_2S_f,S_iS_1}(\omega_1)f_{K_1}(\omega_1)
\Gamma^{K_2}_{S_iS_2,S_1S_f}(\omega_2)f_{K_2}(\omega_2)}
{(E_{S_1}-E_{S_i}-\omega_1+i0)(E_{S_f}-E_{S_i}-\omega_1-\omega_2+i0)(E_{S_i}-E_{S_2}+\omega_2+i0)}
\\
&+
\frac{
\Gamma^{K_1}_{S_2S_f,S_iS_1}(\omega_1)f_{K_1}(\omega_1)
\Gamma^{K_2}_{S_fS_1,S_2S_i}(\omega_2)[1-f_{K_2}(\omega_2)]}
{(E_{S_1}-E_{S_i}-\omega_1+i0)(E_{S_f}-E_{S_i}-\omega_1+\omega_2+i0)(E_{S_i}-E_{S_2}-\omega_2+i0)}
\bigg]
\end{align*}


\subsection{B.(0).(s) projections}
This contribution is of the type
$
 W^{(4)}_{S_i\leftarrow S_i}\, P_{S_i}
$

\subsubsection{Diagram (b)}
\begin{align*}
&2\,\mbox{Im}\sum_{S_1,S_2,S_3}\sum_{K_1,K_2\in\{L,R\}}
\int_{-\infty}^t dt'\int_{-\infty}^{t'}dt_2\int_{-\infty}^{t_2} dt_1 
\\
&\bigg(\ 
\sigma^{K_1\, >}_{S_1S_i,S_1S_2}(t-t')\, g^{<}_{S_1S_2}(t'-t)\, g^{>}_{S_iS_3}(t-t_1)\,
\langle\hat F_{S_iS_3,S_iS_3}(t_1)\rangle \sigma^{K_2\, <}_{S_iS_3,S_2S_3}(t_1-t_2)\,
g^{<}_{S_2S_3}(t_2-t)
\\
&-
\sigma^{K_1\, <}_{S_iS_1,S_2S_1}(t-t')\, g^{>}_{S_2S_1}(t'-t)\,
g^{<}_{S_3S_i}(t-t_1)\,\langle\hat F_{S_3S_i,S_3S_i}(t_1)\rangle\,
\sigma^{K_2\, >}_{S_3S_i,S_3S_2}(t_1-t_2)\, g^{>}_{S_3S_2}(t_2-t)\,
\bigg)
\end{align*}

\subsubsection{Diagram (c)}
\begin{align*}
&2\,\mbox{Im}\sum_{S_1,S_2,S_3}\sum_{K_1,K_2\in\{L,R\}}
\int_{-\infty}^t dt'\int_{-\infty}^{t'}dt_2\int_{-\infty}^{t_2} dt_1 
\\
&\bigg(\ 
\sigma^{K_1\, <}_{S_iS_1,S_2S_1}(t-t')\, g^{<}_{S_2S_1}(t'-t_2)\, g^{<}_{S_3S_i}(t_2-t_1)\,
\sigma^{K_2\, >}_{S_3S_i,S_3S_2}(t_1-t_2)\, g^{>}_{S_iS_1}(t_2-t)\,
\langle\hat F_{S_iS_1,S_iS_1}(t)\rangle
\\
&-
\sigma^{K_1\, >}_{S_1S_i,S_1S_2}(t-t')\, g^{>}_{S_1S_2}(t'-t_2)\, g^{>}_{S_iS_3}(t_2-t_1)\,
\sigma^{K_2\, <}_{S_iS_3,S_2S_3}(t_1-t_2)\, g^{<}_{S_1S_i}(t_2-t)\,
\langle\hat F_{S_1S_i,S_1S_i}(t)\rangle
\bigg)
\end{align*}

\subsubsection{Diagram (d)}
\begin{align*}
& 2\,\mbox{Im}\sum_{S_1,S_2,S_3}\sum_{K_1,K_2\in\{L,R\}}
\int_{-\infty}^t dt'\int_{-\infty}^{t'}dt_2\int_{-\infty}^{t_2} dt_1
\\
&\bigg(\ 
\sigma^{K_1\, <}_{S_iS_1,S_2S_1}(t-t')\, g^{<}_{S_2S_1}(t'-t_2)\,
\sigma^{K_2\, <}_{S_2S_3,S_iS_3}(t_2-t_1)\, g^{>}_{S_iS_3}(t_1-t_2)\,
g^{>}_{S_iS_1}(t_2-t)\,\langle\hat F_{S_iS_1,S_iS_1}(t)\rangle
\\
&
-\sigma^{K_1\, >}_{S_1S_i,S_1S_2}(t-t')\, g^{>}_{S_1S_2}(t'-t_2)\,
\sigma^{K_2\, >}_{S_3S_2,S_3S_i}(t_2-t_1)\, g^{<}_{S_3S_i}(t_1-t_2)\,
g^{<}_{S_1S_i}(t_2-t)\,\langle\hat F_{S_1S_i,S_1S_i}(t)\rangle
\bigg)
\end{align*}

\subsubsection{Diagram (i)}
\begin{align*}
&-2\,\mbox{Im}\sum_{S_1,S_2,S_3}\sum_{K_1,K_2\in\{L,R\}}
\int_{-\infty}^t dt'\int_{-\infty}^{t'}dt_2\int_{-\infty}^{t_2} dt_1 \,
\\
&\bigg(\ 
\sigma^{K_1\, >}_{S_1S_i,S_1S_2}(t-t')\, g^{<}_{S_1S_2}(t'-t)\, B^{>}_{S_iS_2}(t-t_2)\,
\sigma^{K_2\, >}_{S_3S_2,S_3S_i}(t_2-t_1)\, g^{<}_{S_3S_i}(t_1-t_2)
\\
&+
\sigma^{K_1\, <}_{S_iS_1,S_2S_1}(t-t')\, g^{>}_{S_2S_1}(t'-t)\, B^{<}_{S_2S_i}(t-t_2)\,
\sigma^{K_2\, <}_{S_2S_3,S_iS_3}(t_2-t_1)\, g^{>}_{S_iS_3}(t_1-t_2)
\\
& +
\sigma^{K_1\, <}_{S_iS_1,S_2S_1}(t-t')\, g^{>}_{S_2S_1}(t'-t)\, B^{<}_{S_2S_i}(t-t_1)\,
\sigma^{K_2\, >}_{S_3S_i,S_3S_2}(t_1-t_2)\, g^{<}_{S_3S_2}(t_2-t_1)
\\
& +
\sigma^{K_1\, >}_{S_1S_i,S_1S_2}(t-t')\, g^{<}_{S_1S_2}(t'-t)\, B^{>}_{S_iS_2}(t-t_1)\,
\sigma^{K_2\, <}_{S_iS_3,S_2S_3}(t_1-t_2)\, g^{>}_{S_2S_3}(t_2-t_1)
\bigg)
\end{align*}

\subsubsection{Resulting zero-order expression}
\begin{align*}
&2\,\mbox{Im}\sum_{S_1,S_2,S_3}\sum_{K_1,K_2\in\{L,R\}}
\int\frac{d\omega_1}{2\pi}\int\frac{d\omega_2}{2\pi}\,
\\
&\bigg[\quad\bigg(
\frac{
\Gamma^{K_1}_{S_iS_1,S_2S_1}(\omega_1) f_{K_1}(\omega_1)
\Gamma^{K_2}_{S_3S_i,S_3S_2}(\omega_2)[1-f_{K_2}(\omega_2)]}
{(E_{S_3}-E_{S_i}+\omega_2+i0)(E_{S_2}-E_{S_i}+i0)(E_{S_1}-E_{S_i}-\omega_1+i0)}
\\
&\quad +
\frac{
\Gamma^{K_1}_{S_1S_i,S_1S_2}(\omega_1) [1-f_{K_1}(\omega_1)]
\Gamma^{K_2}_{S_iS_3,S_2S_3}(\omega_2) f_{K_2}(\omega_2)}
{(E_{S_i}-E_{S_3}+\omega_2+i0)(E_{S_i}-E_{S_2}+i0)(E_{S_i}-E_{S_1}-\omega_1+i0)}
\bigg)
\\
&\quad\times \bigg(P_{S_2}+\frac{P_{S_i}P_{S_1}}{P_{S_i}+P_{S_3}}\bigg)\frac{P_{S_i}}{P_{S_1}+P_{S_2}}
\\
& +\quad \bigg(
\frac{
\Gamma^{K_1}_{S_iS_1,S_2S_1}(\omega_1) f_{K_1}(\omega_1)
\Gamma^{K_2}_{S_2S_3,S_iS_3}(\omega_2)f_{K_2}(\omega_2)}
{(E_{S_3}-E_{S_i}-\omega_2+i0)(E_{S_2}-E_{S_i}+i0)(E_{S_1}-E_{S_i}-\omega_1+i0)}
\\
&\quad +
\frac{
\Gamma^{K_1}_{S_1S_i,S_1S_2}(\omega_1) [1-f_{K_1}(\omega_1)]
\Gamma^{K_2}_{S_3S_2,S_3S_i}(\omega_2) [1-f_{K_2}(\omega_2)]}
{(E_{S_i}-E_{S_3}-\omega_2+i0)(E_{S_i}-E_{S_2}+i0)(E_{S_i}-E_{S_1}-\omega_1+i0)}
\bigg)
\\
&\quad\times\frac{P_{S_i}^2}{P_{S_i}+P_{S_3}}
\bigg]
\end{align*}


\subsection{B.(1).(s) projections}
This contribution is of the type
$
 W^{(4)}_{S_f\leftarrow S_i}\, P_{S_i}
$

\subsubsection{Diagram (b)}
\begin{align*}
& 2\,\mbox{Im} \sum_{S_1,S_2}\sum_{K_1,K_2\in\{L,R\}}
\int_{-\infty}^t dt' \int_{-\infty}^{t'} dt_2 \int_{-\infty}^{t_2} dt_1
\\
& \bigg(\ 
\sigma^{K_1\, <}_{S_iS_f,S_2S_f}(t-t')\, g^{>}_{S_2S_f}(t'-t)\,
g^{<}_{S_1S_i}(t-t_1)\,\langle\hat F_{S_1S_i,S_1S_i}(t_1)\rangle\,
\sigma^{K_2\, >}_{S_1S_i,S_1S_2}(t_1-t_2)\, g^{>}_{S_1S_2}(t_2-t)
\\
&-
\sigma^{K_1\, >}_{S_fS_i,S_fS_2}(t-t')\, g^{<}_{S_fS_2}(t'-t)\,
g^{>}_{S_iS_1}(t-t_1)\, \langle\hat F_{S_iS_1,S_iS_1}(t_1)\rangle\,
\sigma^{K_2\, <}_{S_iS_1,S_2S_1}(t_1-t_2)\, g^{<}_{S_2S_1}(t_2-t)
\bigg)
\end{align*}

\subsubsection{Diagram (c)}
\begin{align*}
& 2\,\mbox{Im} \sum_{S_1,S_2}\sum_{K_1,K_2\in\{L,R\}}
\int_{-\infty}^t dt' \int_{-\infty}^{t'} dt_2 \int_{-\infty}^{t_2} dt_1
\\
& \bigg(\ 
\sigma^{K_1\, >}_{S_fS_i,S_fS_2}(t-t')\, g^{>}_{S_fS_2}(t'-t_2)\,
g^{>}_{S_iS_1}(t_2-t_1)\,\sigma^{K_2\, <}_{S_iS_1,S_2S_1}(t_1-t_2)\,
g^{<}_{S_fS_i}(t_2-t)\langle\hat F_{S_fS_i,S_fS_i}(t)\rangle
\\
& -
\sigma^{K_1\, <}_{S_iS_f,S_2S_f}(t-t')\, g^{<}_{S_2S_f}(t'-t_2)\,
g^{<}_{S_1S_i}(t_2-t_1)\, \sigma^{K_2\, >}_{S_1S_i,S_1S_2}(t_1-t_2)\,
g^{>}_{S_iS_f}(t_2-t)\langle\hat F_{S_iS_f,S_iS_f}(t)\rangle
\bigg)
\end{align*}

\subsubsection{Diagram (d)}
\begin{align*}
& 2\,\mbox{Im} \sum_{S_1,S_2}\sum_{K_1,K_2\in\{L,R\}}
\int_{-\infty}^t dt' \int_{-\infty}^{t'} dt_2 \int_{-\infty}^{t_2} dt_1
\\
& \bigg(\ 
\sigma^{K_1\, >}_{S_fS_i,S_fS_2}(t-t')\, g^{>}_{S_fS_2}(t'-t_2)\, \sigma^{K_2\, >}_{S_1S_2,S_1S_i}(t_2-t_1)\,
g^{<}_{S_1S_i}(t_1-t_2)\, g^{<}_{S_fS_i}(t_2-t)\,\langle\hat F_{S_fS_i,S_fS_i}(t)\rangle
\\
& -
\sigma^{K_1\, <}_{S_iS_f,S_2S_f}(t-t')\, g^{<}_{S_2S_f}(t'-t_2)\, 
\sigma^{K_2\, <}_{S_2S_1,S_iS_1}(t_2-t_1)\, g^{>}_{S_iS_1}(t_1-t_2)\,
g^{>}_{S_iS_f}(t_2-t)\langle\hat F_{S_iS_f,S_iS_f}(t)\rangle
\bigg)
\end{align*}

\subsubsection{Diagram (i)}
\begin{align*}
&  2\,\mbox{Im} \sum_{S_1,S_2}\sum_{K_1,K_2\in\{L,R\}}
\int_{-\infty}^t dt' \int_{-\infty}^{t'} dt_2 \int_{-\infty}^{t_2} dt_1
\\
& \bigg(\ 
\sigma^{K_1\, >}_{S_fS_i,S_fS_2}(t-t')\, g^{<}_{S_fS_2}(t'-t)\, B^{>}_{S_iS_2}(t-t_1)\,
\sigma^{K_2\, <}_{S_iS_1,S_2S_1}(t_1-t_2)\, g^{>}_{S_2S_1}(t_2-t_1)
\\
& +
\sigma^{K_1\, <}_{S_iS_f,S_2S_f}(t-t')\, g^{>}_{S_2S_f}(t'-t)\, B^{<}_{S_2S_i}(t-t_1)\,
\sigma^{K_2\, >}_{S_1S_i,S_1S_2}(t_1-t_2)\, g^{<}_{S_1S_2}(t_2-t_1)
\\
&+
\sigma^{K_1\, >}_{S_fS_i,S_fS_2}(t-t')\, g^{<}_{S_fS_2}(t'-t)\, B^{>}_{S_iS_2}(t-t_2)\,
\sigma^{K_2\, >}_{S_1S_2,S_1S_i}(t_2-t_1)\, g^{<}_{S_1S_i}(t_1-t_2)
\\
&+
\sigma^{K_1\, <}_{S_iS_f,S_2S_f}(t-t')\, g^{>}_{S_2S_f}(t'-t)\, B^{<}_{S_2S_i}(t-t_2)\,
\sigma^{K_2\, <}_{S_2S_1,S_iS_1}(t_2-t_1)\, g^{>}_{S_iS_1}(t_1-t_2)
\bigg)
\end{align*}

\subsubsection{Resulting zero-order expression}
\begin{align*}
&-2\,\mbox{Im} \sum_{S_1,S_2}\sum_{K_1,K_2\in\{L,R\}}
\int\frac{d\omega_1}{2\pi}\int\frac{d\omega_2}{2\pi}
\\
& \bigg[\quad\bigg(
\frac{
\Gamma^{K_1}_{S_fS_i,S_fS_2}(\omega_1)[1-f_{K_1}(\omega_1)]
\Gamma^{K_2}_{S_1S_2,S_1S_i}(\omega_2)[1-f_{K_2}(\omega_2)]}
{(E_{S_i}-E_{S_1}-\omega_2+i0)(E_{S_i}-E_{S_2}+i0)(E_{S_i}-E_{S_f}-\omega_1+i0)}
\\
& \quad+
\frac{
\Gamma^{K_1}_{S_iS_f,S_2S_f}(\omega_1)f_{K_1}(\omega_1)
\Gamma^{K_2}_{S_2S_1,S_iS_1}(\omega_2)f_{K_2}(\omega_2)}
{(E_{S_1}-E_{S_i}-\omega_2+i0)(E_{S_2}-E_{S_i}+i0)(E_{S_f}-E_{S_i}-\omega_1+i0)}
\bigg)
\\
&\quad\times \frac{P_{S_i}^2}{P_{S_i}+P_{S_1}}
\\
&  +\bigg(
\frac{
\Gamma^{K_1}_{S_fS_i,S_fS_2}(\omega_1)[1-f_{K_1}(\omega_1)]
\Gamma^{K_2}_{S_iS_1,S_2S_1}(\omega_2)f_{K_2}(\omega_2)}
{(E_{S_i}-E_{S_1}+\omega_2+i0)(E_{S_i}-E_{S_2}+i0)(E_{S_i}-E_{S_f}-\omega_1+i0)}
\\
& \quad+
\frac{
\Gamma^{K_1}_{S_iS_f,S_2S_f}(\omega_1)f_{K_1}(\omega_1)
\Gamma^{K_2}_{S_1S_i,S_1S_2}(\omega_2)[1-f_{K_2}(\omega_2)]}
{(E_{S_1}-E_{S_i}+\omega_2+i0)(E_{S_2}-E_{S_i}+i0)(E_{S_f}-E_{S_i}-\omega_1+i0)}
\bigg)
\\
& \quad\times \bigg(P_{S_2}+\frac{P_{S_i}P_{S_f}}{P_{S_i}+P_{S_1}}\bigg)\frac{P_{S_i}}{P_{S_2}+P_{S_f}}
\quad\bigg]
\end{align*}


\subsection{B.(1).(t) projections}
This contribution is of the type
$
 W^{(4)}_{S_f\leftarrow S_i}\, P_{S_i}
$

\subsubsection{Diagram (b)}
\begin{align*}
& 2\,\mbox{Im} \sum_{S_1,S_2}\sum_{K_1,K_2\in\{L,R\}}
\int_{-\infty}^t dt' \int_{-\infty}^t dt_1 \int_{-\infty}^{t_1} dt_2
\\
& \bigg(\ 
\sigma^{K_1\, >}_{S_fS_2,S_fS_i}(t-t')\, g^{<}_{S_fS_i}(t'-t)\, g^{<}_{S_1S_i}(t-t_2)\,
\langle\hat F_{S_1S_i,S_1S_i}(t_2)\rangle\, \sigma^{K_2\, >}_{S_1S_i,S_1S_2}(t_2-t_1)\,
g^{>}_{S_1S_2}(t_1-t)
\\
&-
\sigma^{K_1\, <}_{S_2S_f,S_iS_f}(t-t')\, g^{>}_{S_iS_f}(t'-t)\, g^{>}_{S_iS_1}(t-t_2)\,
\langle \hat F_{S_iS_1,S_iS_1}(t_2)\rangle\, \sigma^{K_2\, <}_{S_iS_1,S_2S_1}(t_2-t_1)\,
g^{<}_{S_2S_1}(t_1-t)
\bigg)
\end{align*}

\subsubsection{Diagram (i)}
\begin{align*}
& 2\,\mbox{Im} \sum_{S_1,S_2}\sum_{K_1,K_2\in\{L,R\}} 
\int_{-\infty}^t dt' \int_{-\infty}^t dt_1 \int_{-\infty}^{t_1} dt_2\,
\\
& \bigg(\ 
\sigma^{K_1\, >}_{S_fS_2,S_fS_i}(t-t')\, g^{<}_{S_fS_i}(t'-t)\, B^{<}_{S_2S_i}(t-t_1)\,
\sigma^{K_2\, <}_{S_2S_1,S_iS_1}(t_1-t_2)\, g^{>}_{S_iS_1}(t_2-t_1)
\\
& +
\sigma^{K_1\, <}_{S_2S_f,S_iS_f}(t-t')\, g^{>}_{S_iS_f}(t'-t)\, B^{>}_{S_iS_2}(t-t_1)\,
\sigma^{K_2\, >}_{S_1S_2,S_1S_i}(t_1-t_2)\, g^{<}_{S_1S_i}(t_2-t_1)
\\
& +
\sigma^{K_1\, >}_{S_fS_2,S_fS_i}(t-t')\, g^{<}_{S_fS_i}(t'-t)\, B^{<}_{S_2S_i}(t-t_2)\,
\sigma^{K_2\, >}_{S_1S_i,S_1S_2}(t_2-t_1)\, g^{<}_{S_1S_2}(t_1-t_2)
\\
&+
\sigma^{K_1\, <}_{S_2S_f,S_iS_f}(t-t')\, g^{>}_{S_iS_f}(t'-t)\, B^{>}_{S_iS_2}(t-t_2)\,
\sigma^{K_2\, <}_{S_iS_1,S_2S_1}(t_2-t_1)\, g^{>}_{S_2S_1}(t_1-t_2)
\bigg)
\end{align*}

\subsubsection{Resulting zero-order expression}
\begin{align*}
&-2\,\mbox{Im}\sum_{S_1,S_2}\sum_{K_1,K_2\in\{L,R\}}
\int\frac{d\omega_1}{2\pi}\int\frac{d\omega_2}{2\pi}
\\
&
\bigg[\quad\bigg(
\frac{
\Gamma^{K_1}_{S_fS_2,S_fS_i}(\omega_1)[1-f_{K_1}(\omega_1)]
\Gamma^{K_2}_{S_2S_1,S_iS_1}(\omega_2) f_{K_2}(\omega_2)}
{(E_{S_1}-E_{S_i}-\omega_2+i0)(E_{S_2}-E_{S_i}+i0)(E_{S_i}-E_{S_f}-\omega_1+i0)}
\\
&\quad+
\frac{
\Gamma^{K_1}_{S_2S_f,S_iS_f}(\omega_1) f_{K_1}(\omega_1)
\Gamma^{K_2}_{S_1S_2,S_1S_i}(\omega_2) [1-f_{K_2}(\omega_2)]}
{(E_{S_i}-E_{S_1}-\omega_2+i0)(E_{S_i}-E_{S_2}+i0)(E_{S_f}-E_{S_i}-\omega_1+i0)}
\bigg)
\\
&\qquad\times\frac{P_{S_i}^3}{(P_{S_i}+P_{S_f})(P_{S_i}+P_{S_1})}
\\
&+\quad\bigg(
\frac{
\Gamma^{K_1}_{S_fS_2,S_fS_i}(\omega_1) [1-f_{K_1}(\omega_1)]
\Gamma^{K_2}_{S_1S_i,S_1S_2}(\omega_2) [1-f_{K_2}(\omega_2)]}
{(E_{S_1}-E_{S_i}+\omega_2+i0)(E_{S_2}-E_{S_i}+i0)(E_{S_i}-E_{S_f}-\omega_1+i0)}
\\
&\quad +
\frac{
\Gamma^{K_1}_{S_2S_f,S_iS_f}(\omega_1) f_{K_1}(\omega_1)
\Gamma^{K_2}_{S_iS_1,S_2S_1}(\omega_2) f_{K_2}(\omega_2)}
{(E_{S_i}-E_{S_1}+\omega_2+i0)(E_{S_i}-E_{S_2}+i0)(E_{S_f}-E_{S_i}-\omega_1+i0)}
\bigg)
\\
& \qquad\times\frac{P_{S_i}^2}{P_{S_i}+P_{S_f}}
\quad\bigg]
\end{align*}

\subsection{B.(2).(t) projections}
This contribution is of the type
$
 W^{(4)}_{S_i\leftarrow S_i}\, P_{S_i}
$

\subsubsection{Diagram (b)}
\begin{align*}
& 2\,\mbox{Im}\sum_{S_1,S_2}\sum_{K_1,K_2\in\{L,R\}}
\int_{-\infty}^{t} dt' \int_{-\infty}^t dt_1\int_{-\infty}^{t_1} dt_2
\\
& \bigg(\ 
\sigma^{K_1\, <}_{S_fS_2,S_iS_2}(t-t')\, g^{>}_{S_iS_2}(t'-t)\, g^{>}_{S_iS_1}(t-t_2)\,
\langle\hat F_{S_iS_1,S_iS_1}(t_2)\rangle\, \sigma^{K_2\, <}_{S_iS_1,S_fS_1}(t_2-t_1)\,
g^{<}_{S_fS_1}(t_1-t)
\\
& -
\sigma^{K_1\, >}_{S_2S_f,S_2S_i}(t-t')\, g^{<}_{S_2S_i}(t'-t)\, g^{<}_{S_1S_i}(t-t_2)\,
\langle\hat F_{S_1S_i,S_1S_i}(t_2)\rangle\, \sigma^{K_2\, >}_{S_1S_i,S_1S_f}(t_2-t_1)\,
g^{>}_{S_1S_f}(t_1-t)
\bigg)
\end{align*}

\subsubsection{Diagram (i)}
\begin{align*}
& -2\,\mbox{Im}\sum_{S_1,S_2}\sum_{K_1,K_2\in\{L,R\}}
\int_{-\infty}^{t} dt' \int_{-\infty}^t dt_1\int_{-\infty}^{t_1} dt_2 
\\
&\bigg(\ 
\sigma^{K_1\, <}_{S_fS_2,S_iS_2}(t-t')\, g^{>}_{S_iS_2}(t'-t)\, B^{>}_{S_iS_f}(t-t_2)\,
\sigma^{K_2\, <}_{S_iS_1,S_fS_1}(t_2-t_1)\, g^{<}_{S_fS_1}(t_1-t_2)
\\
& +
\sigma^{K_1\, >}_{S_2S_f,S_2S_i}(t-t')\, g^{<}_{S_2S_i}(t'-t)\, B^{<}_{S_fS_i}(t-t_2)\,
\sigma^{K_2\, >}_{S_1S_i,S_1S_f}(t_2-t_1)\, g^{<}_{S_1S_f}(t_1-t_2)
\\
& +
\sigma^{K_1\, <}_{S_fS_2,S_iS_2}(t-t')\, g^{>}_{S_iS_2}(t'-t)\, B^{>}_{S_iS_f}(t-t_1)\,
\sigma^{K_2\, >}_{S_1S_f,S_1S_i}(t_1-t_2)\, g^{<}_{S_1S_i}(t_2-t_1)
\\
& +
\sigma^{K_1\, >}_{S_2S_f,S_2S_i}(t-t')\, g^{<}_{S_2S_i}(t'-t)\, B^{<}_{S_fS_i}(t-t_1)\,
\sigma^{K_2\, <}_{S_fS_1,S_iS_1}(t_1-t_2)\, g^{>}_{S_iS_1}(t_2-t_1)
\bigg)
\end{align*}

\subsubsection{Resulting zero-order expression}
\begin{align*}
& 2\,\mbox{Im}\sum_{S_1,S_2}\sum_{K_1,K_2\in\{L,R\}} 
\int\frac{d\omega_1}{2\pi}\int\frac{d\omega_2}{2\pi}
\\
& \bigg[\quad\big(
\frac{
\Gamma^{K_1}_{S_fS_2,S_iS_2}(\omega_1) f_{K_1}(\omega_1)
\Gamma^{K_2}_{S_1S_f,S_1S_i}(\omega_2)[1-f_{K_2}(\omega_2)]
}
{(E_{S_i}-E_{S_1}-\omega_2+i0)(E_{S_i}-E_{S_f}+i0)(E_{S_2}-E_{S_i}-\omega_1+i0)}
\\
& \quad +
\frac{
\Gamma^{K_1}_{S_2S_f,S_2S_i}(\omega_1) [1-f_{K_1}(\omega_1)]
\Gamma^{K_2}_{S_fS_1,S_iS_1}(\omega_2) f_{K_2}(\omega_2)}
{(E_{S_1}-E_{S_i}-\omega_2+i0)(E_{S_f}-E_{S_i}+i0)(E_{S_i}-E_{S_2}-\omega_1+i0)}
\bigg)
\\
&\qquad\times\frac{P_{S_i}^3}{(P_{S_i}+P_{S_1})(P_{S_i}+P_{S_2})}
\\
& \quad +\bigg(
\frac{
\Gamma^{K_1}_{S_fS_2,S_iS_2}(\omega_1) f_{K_1}(\omega_1)
\Gamma^{K_2}_{S_iS_1,S_fS_1}(\omega_2) f_{K_2}(\omega_2)}
{(E_{S_i}-E_{S_1}+\omega_2+i0)(E_{S_i}-E_{S_f}+i0)(E_{S_2}-E_{S_i}-\omega_1+i0)}
\\
& \quad +
\frac{
\Gamma^{K_1}_{S_2S_f,S_2S_i}(\omega_1) [1-f_{K_1}(\omega_1)]
\Gamma^{K_2}_{S_1S_i,S_1S_f}(\omega_2) [1-f_{K_2}(\omega_2)]}
{(E_{S_1}-E_{S_i}+\omega_2+i0)(E_{S_f}-E_{S_i}+i0)(E_{S_i}-E_{S_2}-\omega_1+i0)}
\bigg)
\\
&\qquad\times\frac{P_{S_i}^2}{P_{S_i}+P_{S_2}}
\quad\bigg]
\end{align*}


\subsection{C.(0).(s) projections}
This contribution is of the type
$
 W^{(4)}_{S_i\leftarrow S_i}\, P_{S_i}
$

\subsubsection{Diagram (c)}
\begin{align*}
&2\,\mbox{Im}\sum_{S_1,S_2,S_3}\sum_{K_1,K_2\in\{L,R\}} 
\int_{-\infty}^t dt_1 \int_{-\infty}^{t_1} dt_2 \int_{-\infty}^{t_2} dt'
\\
&\bigg(\ 
\sigma^{K_1\, >}_{S_1S_i,S_3S_i}(t-t')\, g^{<}_{S_3S_i}(t'-t_2)\, g^{<}_{S_2S_1}(t_2-t_1)\,
\sigma^{K_2\, >}_{S_2S_1,S_2S_3}(t_1-t_2)\, g^{<}_{S_1S_i}(t_2-t)\langle\hat F_{S_1S_i,S_1S_i}(t)\rangle
\\
& -
\sigma^{K_1\, <}_{S_iS_1,S_iS_3}(t-t')\, g^{>}_{S_iS_3}(t'-t_2)\, g^{>}_{S_1S_2}(t_2-t_1)\,
\sigma^{K_2\, <}_{S_1S_2,S_3S_2}(t_1-t_2)\, g^{>}_{S_iS_1}(t_2-t)\langle\hat F_{S_iS_1,S_iS_1}(t)\rangle
\bigg)
\end{align*}

\subsubsection{Diagram (d)}
\begin{align*}
& 2\,\mbox{Im}\sum_{S_1,S_2,S_3}\sum_{K_1,K_2\in\{L,R\}} 
\int_{-\infty}^t dt_1 \int_{-\infty}^{t_1} dt_2 \int_{-\infty}^{t_2} dt'
\\
&\bigg(\ 
\sigma^{K_1\, >}_{S_1S_i,S_3S_i}(t-t')\, g^{<}_{S_3S_i}(t'-t_2)\, \sigma^{K_2\, <}_{S_3S_2,S_1S_2}(t_2-t_1)\,
g^{>}_{S_1S_2}(t_1-t_2)\, g^{<}_{S_1S_i}(t_2-t)\langle\hat F_{S_1S_i,S_1S_i}(t)\rangle
\\
& -
\sigma^{K_1\, <}_{S_iS_1,S_iS_3}(t-t')\, g^{>}_{S_iS_3}(t'-t_2)\, \sigma^{K_2\, >}_{S_2S_3,S_2S_1}(t_2-t_1)\,
g^{<}_{S_2S_1}(t_1-t_2)\, g^{>}_{S_iS_1}(t_2-t)\langle\hat F_{S_iS_1,S_iS_1}(t)\rangle
\bigg)
\end{align*}

\subsubsection{Diagram (g)}
\begin{align*}
&2\,\mbox{Im}\sum_{S_1,S_2,S_3}\sum_{K_1,K_2\in\{L,R\}}
\int_{-\infty}^t dt_1 \int_{-\infty}^{t_1} dt_2 \int_{-\infty}^{t_2} dt'
\\
&\bigg(
\sigma^{K_1\, <}_{S_iS_1,S_iS_3}(t-t')\, g^{>}_{S_iS_3}(t'-t_2)\,
 \sigma^{K_2\, >}_{S_2S_3,S_2S_1}(t_2-t_1)\, g^{>}_{S_2S_1}(t_1-t)\,
 B^{<}_{S_2S_i}(t-t_2)
 \\
 &+
 \sigma^{K_1\, >}_{S_1S_i,S_3S_i}(t-t')\, g^{<}_{S_3S_i}(t'-t_2)\,
 \sigma^{K_2\, <}_{S_3S_2,S_1S_2}(t_2-t_1)\, g^{<}_{S_1S_2}(t_1-t)\,
 B^{>}_{S_iS_2}(t-t_2)
 \bigg)
\end{align*}

\subsubsection{Diagram (h)}
\begin{align*}
&2\,\mbox{Im}\sum_{S_1,S_2,S_3}\sum_{K_1,K_2\in\{L,R\}}
\int_{-\infty}^t dt_1 \int_{-\infty}^{t_1} dt_2 \int_{-\infty}^{t_2} dt'
\\
&\bigg(\ 
\sigma^{K_1\, >}_{S_1S_i,S_3S_i}(t-t')\, g^{<}_{S_3S_i}(t'-t_2)\, D^{<}_{S_2S_i}(t_2-t)\,
g^{>}_{S_2S_1}(t-t_1)\, \sigma^{K_2\, >}_{S_2S_1,S_2S_3}(t_1-t_2)
\\
&+
\sigma^{K_1\, <}_{S_iS_1,S_iS_3}(t-t')\, g^{>}_{S_iS_3}(t'-t_2)\, D^{>}_{S_iS_2}(t_2-t)\,
g^{<}_{S_1S_2}(t-t_1)\, \sigma^{K_2\, <}_{S_1S_2,S_3S_2}(t_1-t_2)
\bigg)
\end{align*}

\subsubsection{Resulting zero-order expression}
\begin{align*}
&2\,\mbox{Im}\sum_{S_1,S_2,S_3}\sum_{K_1,K_2\in\{L,R\}}
\int\frac{d\omega_1}{2\pi} \int\frac{d\omega_2}{2\pi}\,
\frac{P_{S_i}^2}{P_{S_i}+P_{S_3}}
\\
&\bigg[\ 
\frac{
\Gamma^{K_1}_{S_iS_1,S_iS_3}(\omega_1) f_{K_1}(\omega_1)
\Gamma^{K_2}_{S_2S_3,S_2S_1}(\omega_2) [1-f_{K_2}(\omega_2)]
}
{(E_{S_3}-E_{S_i}-\omega_1+i0)(E_{S_2}-E_{S_i}-\omega_1+\omega_2+i0)(E_{S_1}-E_{S_i}-\omega_1+i0)}
\\
&+
\frac{
\Gamma^{K_1}_{S_iS_1,S_iS_3}(\omega_1) f_{K_1}(\omega_1)
\Gamma^{K_2}_{S_1S_2,S_3S_2}(\omega_2) f_{K_2}(\omega_2)
}
{(E_{S_3}-E_{S_i}-\omega_1+i0)(E_{S_2}-E_{S_i}-\omega_1-\omega_2+i0)(E_{S_1}-E_{S_i}-\omega_1+i0)}
\\
&+
\frac{
\Gamma^{K_1}_{S_1S_i,S_3S_i}(\omega_1) [1-f_{K_1}(\omega_1)]
\Gamma^{K_2}_{S_2S_1,S_2S_3}(\omega_2) [1-f_{K_2}(\omega_2)]
}
{(E_{S_i}-E_{S_3}-\omega_1+i0)(E_{S_i}-E_{S_2}-\omega_1-\omega_2+i0)(E_{S_i}-E_{S_1}-\omega_1+i0)}
\\
&+
\frac{
\Gamma^{K_1}_{S_1S_i,S_3S_i}(\omega_1) [1-f_{K_1}(\omega_1)]
\Gamma^{K_2}_{S_3S_2,S_1S_2}(\omega_2) f_{K_2}(\omega_2)
}
{(E_{S_i}-E_{S_3}-\omega_1+i0)(E_{S_i}-E_{S_2}-\omega_1+\omega_2+i0)(E_{S_i}-E_{S_1}-\omega_1+i0)}
\bigg]
\end{align*}


\subsection{C.(1).(s) projections}
This contribution is of the type
$
 W^{(4)}_{S_f\leftarrow S_i}\, P_{S_i}
$

\subsubsection{Diagram (c)}
\begin{align*}
& 2\,\mbox{Im}\sum_{S_1,S_2}\sum_{K_1,K_2\in\{L,R\}} 
\int_{-\infty}^{t} dt_1 \int_{-\infty}^{t_1} dt_2 \int_{-\infty}^{t_2} dt'
\\
&  \bigg(\ 
\sigma^{K_1\, <}_{S_iS_f,S_iS_1}(t-t')\, g^{>}_{S_iS_1}(t'-t_2)\, g^{>}_{S_fS_2}(t_2-t_1)\,
\sigma^{K_2\, <}_{S_fS_2,S_1S_2}(t_1-t_2)\, g^{>}_{S_iS_f}(t_2-t)
\langle\hat F_{S_iS_f,S_iS_f}(t)\rangle
\\
& -
\sigma^{K_1\, >}_{S_fS_i,S_1S_i}(t-t')\, g^{<}_{S_1S_i}(t'-t_2)\, g^{<}_{S_2S_f}(t_2-t_1)\,
\sigma^{K_2\, >}_{S_2S_f,S_2S_1}(t_1-t_2)\, g^{<}_{S_fS_i}(t_2-t)
\langle\hat F_{S_fS_i,S_fS_i}(t)\rangle
\bigg)
\end{align*}

\subsubsection{Diagram (d)}
\begin{align*}
& 2\,\mbox{Im}\sum_{S_1,S_2}\sum_{K_1,K_2\in\{L,R\}} 
\int_{-\infty}^{t} dt_1 \int_{-\infty}^{t_1} dt_2 \int_{-\infty}^{t_1} dt'
\\
&\bigg(\ 
\sigma^{K_1\, <}_{S_iS_f,S_iS_1}(t-t')\, g^{>}_{S_iS_1}(t'-t_2)\,
\sigma^{K_2\, >}_{S_2S_1,S_2S_f}(t_2-t_1)\,  g^{<}_{S_2S_f}(t_1-t_2)\,
g^{>}_{S_iS_f}(t_2-t)\, \langle\hat F_{S_iS_f,S_iS_f}(t)\rangle
\\
& -
\sigma^{K_1\, >}_{S_fS_i,S_1S_i}(t-t')\, g^{<}_{S_1S_i}(t'-t_2)\,
\sigma^{K_2\, <}_{S_1S_2,S_fS_2}(t_2-t_1)\,  g^{>}_{S_fS_2}(t_1-t_2)\,
g^{<}_{S_fS_i}(t_2-t)\, \langle\hat F_{S_fS_i,S_fS_i}(t)\rangle
\bigg)
\end{align*}

\subsubsection{Diagram (g)}
\begin{align*}
& -2\,\mbox{Im}\sum_{S_1,S_2}\sum_{K_1,K_2\in\{L,R\}}
\int_{-\infty}^{t} dt_1 \int_{-\infty}^{t_1} dt_2 \int_{-\infty}^{t_2} dt'
\\
& \bigg(\ 
\sigma^{K_1\, >}_{S_fS_i,S_1S_i}(t-t')\, g^{<}_{S_1S_i}(t'-t_2)\, 
\sigma^{K_2\, <}_{S_1S_2,S_fS_2}(t_2-t_1)\,  g^{<}_{S_fS_2}(t_1-t)\,
B^{>}_{S_iS_2}(t-t_2)
\\
&+ 
\sigma^{K_1\, <}_{S_iS_f,S_iS_1}(t-t')\, g^{>}_{S_iS_1}(t'-t_2)\, 
\sigma^{K_2\, >}_{S_2S_1,S_2S_f}(t_2-t_1)\,  g^{>}_{S_2S_f}(t_1-t)\,
B^{<}_{S_2S_i}(t-t_2)
\bigg)
\end{align*}

\subsubsection{Diagram (h)}
\begin{align*}
& -2\,\mbox{Im}\sum_{S_1,S_2}\sum_{K_1,K_2\in\{L,R\}} 
\int_{-\infty}^{t} dt_1 \int_{-\infty}^{t_1} dt_2 \int_{-\infty}^{t_2} dt'
\\
& \bigg(\ 
\sigma^{K_1\, >}_{S_fS_i,S_1S_i}(t-t')\, g^{<}_{S_1S_i}(t'-t_2)\, D^{<}_{S_2S_i}(t_2-t)\,
g^{>}_{S_2S_f}(t-t_1)\,\sigma^{K_2\, >}_{S_2S_f,S_2S_1}(t_1-t_2)
\\
& +
\sigma^{K_1\, <}_{S_iS_f,S_iS_1}(t-t')\, g^{>}_{S_iS_1}(t'-t_2)\, D^{>}_{S_iS_2}(t_2-t)\,
g^{<}_{S_fS_2}(t-t_1)\,\sigma^{K_2\, <}_{S_fS_2,S_1S_2}(t_1-t_2)
\bigg)
\end{align*}

\subsubsection{Resulting zero-order expression}
\begin{align*}
& -2\,\mbox{Im}\sum_{S_1,S_2}\sum_{K_1,K_2\in\{L,R\}} 
\int\frac{d\omega_1}{2\pi} \int\frac{d\omega_2}{2\pi}\, \frac{P_{S_i}^2}{P_{S_i}+P_{S_1}}
\\
&\bigg[\ 
\frac{
\Gamma^{K_1}_{S_fS_i,S_1S_i}(\omega_1) [1-f_{K_1}(\omega_1)]
\Gamma^{K_2}_{S_2S_f,S_2S_1}(\omega_2) [1-f_{K_2}(\omega_2)]
}
{(E_{S_i}-E_{S_1}-\omega_1+i0)(E_{S_i}-E_{S_2}-\omega_1-\omega_2+i0)(E_{S_i}-E_{S_f}-\omega_1+i0)}
\\
& +
\frac{
\Gamma^{K_1}_{S_fS_i,S_1S_i}(\omega_1) [1-f_{K_1}(\omega_1)]
\Gamma^{K_2}_{S_1S_2,S_fS_2}(\omega_2) f_{K_2}(\omega_2)
}
{(E_{S_i}-E_{S_1}-\omega_1+i0)(E_{S_i}-E_{S_2}-\omega_1+\omega_2+i0)(E_{S_i}-E_{S_f}-\omega_1+i0)}
\\
& +
\frac{
\Gamma^{K_1}_{S_iS_f,S_iS_1}(\omega_1) f_{K_1}(\omega_1)
\Gamma^{K_2}_{S_2S_1,S_2S_f}(\omega_2) [1-f_{K_2}(\omega_2)]
}
{(E_{S_1}-E_{S_i}-\omega_1+i0)(E_{S_2}-E_{S_i}-\omega_1+\omega_2+i0)(E_{S_f}-E_{S_i}-\omega_1+i0)}
\\
& +
\frac{
\Gamma^{K_1}_{S_iS_f,S_iS_1}(\omega_1) f_{K_1}(\omega_1)
\Gamma^{K_2}_{S_fS_2,S_1S_2}(\omega_2) f_{K_2}(\omega_2)
}
{(E_{S_1}-E_{S_i}-\omega_1+i0)(E_{S_2}-E_{S_i}-\omega_1-\omega_2+i0)(E_{S_f}-E_{S_i}-\omega_1+i0)}
\bigg]
\end{align*}


\subsection{C.(1).(t) projections}
This contribution is of the type
$
 W^{(4)}_{S_f\leftarrow S_i}\, P_{S_i}
$

\subsubsection{Diagram (b)}
\begin{align*}
& 2\,\mbox{Im}\sum_{S_1,S_2}\sum_{K_1,K_2\in\{L,R\}}
\int_{-\infty}^t dt' \int_{-\infty}^{t'} dt_2 \int_{-\infty}^{t} dt_1\,
\\
& \bigg(\ 
\sigma^{K_1\, >}_{S_2S_f,S_2S_1}(t-t')\, g^{<}_{S_2S_1}(t'-t)\, g^{>}_{S_iS_1}(t-t_2)\,
\langle\hat F_{S_iS_1,S_iS_1}(t_2)\rangle\, 
\sigma^{K_2\, <}_{S_iS_1,S_iS_f}(t_2-t_1)\, g^{>}_{S_iS_f}(t_1-t)
\\
& -
\sigma^{K_1\, <}_{S_fS_2,S_1S_2}(t-t')\, g^{>}_{S_1S_2}(t'-t)\, g^{<}_{S_1S_i}(t-t_2)\,
\langle\hat F_{S_1S_i,S_1S_i}(t_2)\rangle\, 
\sigma^{K_2\, >}_{S_1S_i,S_fS_i}(t_2-t_1)\, g^{<}_{S_fS_i}(t_1-t)
\\
& +
\sigma^{K_1\, <}_{S_fS_2,S_1S_2}(t-t')\, g^{>}_{S_1S_2}(t'-t)\, g^{>}_{S_iS_f}(t-t_1)\,
\langle\hat F_{S_iS_f,S_iS_f}(t_1)\rangle\, 
\sigma^{K_2\, <}_{S_iS_f,S_iS_1}(t_1-t_2)\, g^{>}_{S_iS_1}(t_2-t)
\\
& -
\sigma^{K_1\, >}_{S_2S_f,S_2S_1}(t-t')\, g^{<}_{S_2S_1}(t'-t)\, g^{<}_{S_fS_i}(t-t_1)\,
\langle\hat F_{S_fS_i,S_fS_i}(t_1)\rangle\, 
\sigma^{K_2\, >}_{S_fS_i,S_1S_i}(t_1-t_2)\, g^{<}_{S_1S_i}(t_2-t)
\bigg)
\end{align*}

\subsubsection{Diagram (g)}
\begin{align*}
 & -2\,\mbox{Im}\sum_{S_1,S_2}\sum_{K_1,K_2\in\{L,R\}} 
\int_{-\infty}^t dt' \int_{-\infty}^{t'} dt_2 \int_{-\infty}^{t} dt_1
\\
& \bigg(\ 
\sigma^{K_1\, <}_{S_fS_2,S_1S_2}(t-t')\, g^{<}_{S_1S_2}(t'-t_2)\, 
\sigma^{K_2\, >}_{S_1S_i,S_fS_i}(t_2-t_1)\, g^{<}_{S_fS_i}(t_1-t)\, B^{<}_{S_2S_i}(t-t_2)
\\
& +
\sigma^{K_1\, >}_{S_2S_f,S_2S_1}(t-t')\, g^{>}_{S_2S_1}(t'-t_2)\, 
\sigma^{K_2\, <}_{S_iS_1,S_iS_f}(t_2-t_1)\, g^{>}_{S_iS_f}(t_1-t)\, B^{>}_{S_iS_2}(t-t_2)
\bigg)
\end{align*}

\subsubsection{Diagram (h)}
\begin{align*}
& -2\,\mbox{Im}\sum_{S_1,S_2}\sum_{K_1,K_2\in\{L,R\}}
\int_{-\infty}^t dt' \int_{-\infty}^{t'} dt_2 \int_{-\infty}^{t} dt_1
\\
& \bigg(\ 
\sigma^{K_1\, <}_{S_fS_2,S_1S_2}(t-t')\, g^{<}_{S_1S_2}(t'-t_2)\, D^{>}_{S_iS_2}(t_2-t)\,
g^{>}_{S_iS_f}(t-t_1)\, \sigma^{K_2\, <}_{S_iS_f,S_iS_1}(t_1-t_2)
\\
& +
\sigma^{K_1\, >}_{S_2S_f,S_2S_1}(t-t')\, g^{>}_{S_2S_1}(t'-t_2)\, D^{<}_{S_2S_i}(t_2-t)\,
g^{<}_{S_fS_i}(t-t_1)\, \sigma^{K_2\, >}_{S_fS_i,S_1S_i}(t_1-t_2)
\bigg)
\end{align*}

\subsubsection{Resulting zero-order expression}
\begin{align*}
& -2\,\mbox{Im} \sum_{S_1,S_2}\sum_{K_1,K_2\in\{L,R\}}
\int\frac{d\omega_1}{2\pi} \int\frac{d\omega_2}{2\pi}
\\
& \bigg[\quad\bigg(
\frac{
\Gamma^{K_1}_{S_fS_2,S_1S_2}(\omega_1) f_{K_1}(\omega_1)
\Gamma^{K_2}_{S_1S_i,S_fS_i}(\omega_2) [1-f_{K_2}(\omega_2)]
}
{(E_{S_1}-E_{S_i}+\omega_2+i0)(E_{S_2}-E_{S_i}-\omega_1+\omega_2+i0)(E_{S_i}-E_{S_f}-\omega_2+i0)}
\\
& \quad+
\frac{
\Gamma^{K_1}_{S_2S_f,S_2S_1}(\omega_1) [1-f_{K_1}(\omega_1)]
\Gamma^{K_2}_{S_iS_1,S_iS_f}(\omega_2) f_{K_2}(\omega_2)
}
{(E_{S_i}-E_{S_1}+\omega_2+i0)(E_{S_i}-E_{S_2}-\omega_1+\omega_2+i0)(E_{S_f}-E_{S_i}-\omega_2+i0)}
\bigg)
\\
& \quad\times\frac{P_{S_i}^2}{P_{S_i}+P_{S_f}}
\\
& \quad+\bigg(
\frac{
\Gamma^{K_1}_{S_fS_2,S_1S_2}(\omega_1) f_{K_1}(\omega_1)
\Gamma^{K_2}_{S_iS_f,S_iS_1}(\omega_2) f_{K_2}(\omega_2)
}
{(E_{S_1}-E_{S_i}-\omega_2+i0)(E_{S_2}-E_{S_i}-\omega_1-\omega_2+i0)(E_{S_i}-E_{S_f}+\omega_2+i0)}
\\
& \quad+
\frac{
\Gamma^{K_1}_{S_2S_f,S_2S_1}(\omega_1) [1-f_{K_1}(\omega_1)]
\Gamma^{K_2}_{S_fS_i,S_1S_i}(\omega_2) [1-f_{K_2}(\omega_2)]
}
{(E_{S_i}-E_{S_1}-\omega_2+i0)(E_{S_i}-E_{S_2}-\omega_1-\omega_2+i0)(E_{S_f}-E_{S_i}+\omega_2+i0)}
\big)
\\
&
\quad\times \bigg(\frac{P_{S_2}}{P_{S_i}+P_{S_f}}+\frac{P_{S_1}}{P_{S_i}+P_{S_1}}\bigg) \frac{P_{S_i}^2}{P_{S_1}+P_{S_2}}
\quad\bigg]
\end{align*}


\subsection{C.(2).(t) projections}
This contribution is of the type
$
 W^{(4)}_{S_f\leftarrow S_i}\, P_{S_i}
$

\subsubsection{Diagram (b)}
\begin{align*}
& 2\,\mbox{Im}\sum_{S_1,S_2}\sum_{K_1,K_2\in\{L,R\}}
\int_{-\infty}^{t} dt' \int_{-\infty}^{t'} dt_2 \int_{-\infty}^{t} dt_1\,
\\
& \bigg(\ 
\sigma^{K_1\, >}_{S_fS_2,S_fS_1}(t-t')\, g^{<}_{S_fS_1}(t'-t)\, g^{<}_{S_2S_i}(t-t_1)\,
\langle\hat F_{S_2S_i,S_2S_i}(t_1)\rangle\, 
\sigma^{K_2\, >}_{S_2S_i,S_1S_i}(t_1-t_2)\, g^{<}_{S_1S_i}(t_2-t)
\\
& -
\sigma^{K_1\, <}_{S_2S_f,S_1S_f}(t-t')\, g^{>}_{S_1S_f}(t'-t)\, g^{>}_{S_iS_2}(t-t_1)\,
\langle\hat F_{S_iS_2,S_iS_2}(t_1)\rangle\, 
\sigma^{K_2\, <}_{S_iS_2,S_iS_1}(t_1-t_2)\, g^{>}_{S_iS_1}(t_2-t)
\\
& +
\sigma^{K_1\, <}_{S_2S_f,S_1S_f}(t-t')\, g^{>}_{S_1S_f}(t'-t)\, g^{<}_{S_1S_i}(t-t_2)\,
\langle\hat F_{S_1S_i,S_1S_i}(t_2)\rangle\, 
\sigma^{K_2\, >}_{S_1S_i,S_2S_i}(t_2-t_1)\, g^{<}_{S_2S_i}(t_1-t)
\\
& -
\sigma^{K_1\, >}_{S_fS_2,S_fS_1}(t-t')\, g^{<}_{S_fS_1}(t'-t)\, g^{>}_{S_iS_1}(t-t_2)\,
\langle\hat F_{S_iS_1,S_iS_1}(t_2)\rangle\, 
\sigma^{K_2\, <}_{S_iS_1,S_iS_2}(t_2-t_1)\, g^{>}_{S_iS_2}(t_1-t)
\bigg)
\end{align*}

\subsubsection{Diagram (g)}
\begin{align*}
& 2\,\mbox{Im}\sum_{S_1,S_2}\sum_{K_1,K_2\in\{L,R\}}
\int_{-\infty}^{t} dt' \int_{-\infty}^{t'} dt_2 \int_{-\infty}^{t} dt_1
\\
& \bigg(\ 
\sigma^{K_1\, >}_{S_fS_2,S_fS_1}(t-t')\, g^{>}_{S_fS_1}(t'-t_2)\, 
\sigma^{K_2\, <}_{S_iS_1,S_iS_2}(t_2-t_1)\, g^{>}_{S_iS_2}(t_1-t)\, B^{>}_{S_iS_f}(t-t_2)
\\
& +
\sigma^{K_1\, <}_{S_2S_f,S_1S_f}(t-t')\, g^{<}_{S_1S_f}(t'-t_2)\, 
\sigma^{K_2\, >}_{S_1S_i,S_2S_i}(t_2-t_1)\, g^{<}_{S_2S_i}(t_1-t)\, B^{<}_{S_fS_i}(t-t_2)
\big)
\end{align*}

\subsubsection{Diagram (h)}
\begin{align*}
& 2\,\mbox{Im}\sum_{S_1,S_2}\sum_{K_1,K_2\in\{L,R\}}
\int_{-\infty}^{t} dt' \int_{-\infty}^{t'} dt_2 \int_{-\infty}^{t} dt_1
\\
& \bigg(\ 
\sigma^{K_1\, >}_{S_fS_2,S_fS_1}(t-t')\, g^{>}_{S_fS_1}(t'-t_2)\, D^{<}_{S_fS_i}(t_2-t)\,
g^{<}_{S_2S_i}(t-t_1)\, \sigma^{K_2\, >}_{S_2S_i,S_1S_i}(t_1-t_2)
\\
& +
\sigma^{K_1\, <}_{S_2S_f,S_1S_f}(t-t')\, g^{<}_{S_1S_f}(t'-t_2)\, D^{>}_{S_iS_f}(t_2-t)\,
g^{>}_{S_iS_2}(t-t_1)\, \sigma^{K_2\, <}_{S_iS_2,S_iS_1}(t_1-t_2)
\bigg)
\end{align*}

\subsubsection{Resulting zero-order expression}
\begin{align*}
& 2\,\mbox{Im}\sum_{S_1,S_2}\sum_{K_1,K_2\in\{L,R\}} 
\int\frac{d\omega_1}{2\pi} \int\frac{d\omega_2}{2\pi}
\\
 & \bigg[\quad\bigg(
\frac{
\Gamma^{K_1}_{S_fS_2,S_fS_1}(\omega_1) [1-f_{K_1}(\omega_1)]
\Gamma^{K_2}_{S_iS_1,S_iS_2}(\omega_2) f_{K_2}(\omega_2)
}
{(E_{S_i}-E_{S_1}+\omega_2+i0)(E_{S_i}-E_{S_f}-\omega_1+\omega_2+i0)(E_{S_2}-E_{S_i}-\omega_2+i0)}
\\
& \quad +
\frac{
\Gamma^{K_1}_{S_2S_f,S_1S_f}(\omega_1) f_{K_1}(\omega_1)
\Gamma^{K_2}_{S_1S_i,S_2S_i}(\omega_2) [1-f_{K_2}(\omega_2)]
}
{(E_{S_1}-E_{S_i}+\omega_2+i0)(E_{S_f}-E_{S_i}-\omega_1+\omega_2+i0)(E_{S_i}-E_{S_2}-\omega_2+i0)}
\bigg)
\\
&\quad\times\frac{P_{S_i}^2}{P_{S_i}+P_{S_2}}
\\
&\quad +\bigg(
\frac{
\Gamma^{K_1}_{S_fS_2,S_fS_1}(\omega_1) [1-f_{K_1}(\omega_1)]
\Gamma^{K_2}_{S_2S_i,S_1S_i}(\omega_2) [1-f_{K_2}(\omega_2)]
}
{(E_{S_i}-E_{S_1}-\omega_2+i0)(E_{S_i}-E_{S_f}-\omega_1-\omega_2+i0)(E_{S_2}-E_{S_i}+\omega_2+i0)}
\\
&\quad +
\frac{
\Gamma^{K_1}_{S_2S_f,S_1S_f}(\omega_1) f_{K_1}(\omega_1)
\Gamma^{K_2}_{S_iS_2,S_iS_1}(\omega_2) f_{K_2}(\omega_2)
}
{(E_{S_1}-E_{S_i}-\omega_2+i0)(E_{S_f}-E_{S_i}-\omega_1-\omega_2+i0)(E_{S_i}-E_{S_2}+\omega_2+i0)}
\bigg)
\\
&\quad\times\bigg(\frac{P_{S_1}}{P_{S_i}+P_{S_1}}+\frac{P_{S_f}}{P_{S_i}+P_{S_2}}\bigg)
\frac{P_{S_i}^2}{P_{S_1}+P_{S_f}}
\quad\bigg]
\end{align*}



\providecommand{\latin}[1]{#1}
\makeatletter
\providecommand{\doi}
  {\begingroup\let\do\@makeother\dospecials
  \catcode`\{=1 \catcode`\}=2\doi@aux}
\providecommand{\doi@aux}[1]{\endgroup\texttt{#1}}
\makeatother
\providecommand*\mcitethebibliography{\thebibliography}
\csname @ifundefined\endcsname{endmcitethebibliography}
  {\let\endmcitethebibliography\endthebibliography}{}